\documentclass[sigconf,natbib=false]{acmart}
\usepackage{float}
\usepackage{graphicx} % for including graphics
\usepackage{subcaption} % for subfigures

\AtBeginDocument{%
  }
\RequirePackage[
  datamodel=acmdatamodel,
  style=acmnumeric,
  ]{biblatex}

\begin{document}

%%
%% The "title" command has an optional parameter,
%% allowing the author to define a "short title" to be used in page headers.
\title{Evaluating the Efficacy of Open-Source LLMs in Enterprise-Specific RAG Systems: A Comparative Study of Performance and Scalability}

%%
%% The "author" command and its associated commands are used to define
%% the authors and their affiliations.
%% Of note is the shared affiliation of the first two authors, and the
%% "authornote" and "authornotemark" commands
%% used to denote shared contribution to the research.
\author{Gautam B}
\authornote{Both authors contributed equally to this research.}
\email{gautamb1a2b@gmail.com}
\orcid{0009-0005-4256-5274}
\affiliation{%
  \institution{Indian Institute of Technology Madras}
  \city{Chennai}
  \state{Tamil Nadu}
  \country{India}
}
%\authornote {Equal contributions by both authors}
\author{Anupam Purwar}
\email{anupam.aiml@gmail.com}
\authornotemark[1]
\orcid{0000-0002-4484-7575}
\affiliation{%
  \institution{Independent}
  \city{Delhi}
  \country{India}
}

\authornote{Corresponding author: anupam.aiml@gmail.com}

%%
%% By default, the full list of authors will be used in the page
%% headers. Often, this list is too long, and will overlap
%% other information printed in the page headers. This command allows
%% the author to define a more concise list
%% of authors' names for this purpose.
\renewcommand{\shortauthors}{Purwar et al.}

%%
%% The abstract is a short summary of the work to be presented in the
%% article.
\begin{abstract}
  This paper presents an analysis of open-source large language models (LLMs) and their application in Retrieval-Augmented Generation (RAG) tasks, specific for enterprise-specific data sets scraped from their websites. With the increasing reliance on LLMs in natural language processing, it is crucial to evaluate their performance, accessibility, and integration within specific organizational contexts. This study examines various open-source LLMs, explores their integration into RAG frameworks using enterprise-specific data, and assesses the performance of different open-source embeddings in enhancing the retrieval and generation process. Our findings indicate that open-source LLMs, combined with effective embedding techniques, can significantly improve the accuracy and efficiency of RAG systems, offering a viable alternative to proprietary solutions for enterprises.
\end{abstract}

%%
%% The code below is generated by the tool at http://dl.acm.org/ccs.cfm.
%% Please copy and paste the code instead of the example below.
%%

\begin{CCSXML}
<ccs2012>
   <concept>
       <concept_id>10010147.10010178.10010179.10010182</concept_id>
       <concept_desc>Computing methodologies~Natural language generation</concept_desc>
       <concept_significance>300</concept_significance>
       </concept>
 </ccs2012>
\end{CCSXML}

\ccsdesc[300]{Computing methodologies~Natural language generation}

%%
%% Keywords. The author(s) should pick words that accurately describe
%% the work being presented. Separate the keywords with commas.
\keywords{Large language models(LLMs), Retrieval Augmented Generation (RAG), Natural language processing(NLP),information retriever, ROUGE score, 
cosine similarity, top-k, Llama3, Mistral,Generative Pre-Trained Transformers(GPT),Generative AI(Gen AI),Cosine Similarity with Groundtruth Answer(CSGA)}

%%
%% This command processes the author and affiliation and title
%% information and builds the first part of the formatted document.
\maketitle

\section{Introduction}
The rapid advancements in natural language processing (NLP) have led to the
development of sophisticated large language models (LLMs) that excel in tasks such as
text generation, summarization, and question answering. Among these advancements, 
Retrieval-Augmented Generation (RAG) has emerged as a promising approach for the retrieval-based systems with generative models to produce highly accurate and contextually relevant 
outputs.The concept of Retrieval-Augmented Generation (RAG) was introduced by Lewis et
al. In their seminar 2020 paper titled "Retrieval-Augmented Generation for Knowledge-
Intensive NLP Tasks."\cite{lewis2020retrieval}.In their research, Lewis et al. present a method that combines 
retrieval-based and generative models to enhance the performance of knowledge-intensive 
tasks. By integrating non-parametric memory (retrieved documents) with parametric 
memory (the generative model's internal parameters), RAG models achieve superior 
accuracy and flexibility in tasks such as open-domain question answering and 
abstract question answering. Karpukhin et al. (2020) developed dense passage 
retrieval for open-domain question answering, which significantly boosts retrieval 
accuracy by using dense vector representations and a neural retriever \cite{karpukhin2020dense}.More recent
work further advances the field by introducing novel 
methodologies for fine-tuning LLMs specifically for RAG tasks in knowledge-intensive 
environments \cite{Afinetune}. There has been efficient ways to improve the retrieval process such 
as the Keyword Augmented Retrieval (KAR), which integrates keyword generation using 
transformer models with document metadata to identify the right context quickly and 
cost-effectively \cite{purwar2024keyword}. Also, approach to handle sparse information where classical RAG using hybrid retriever fails to generate correct answers have been reported \cite{cos}. More recent work by Tay et al. (2023) on the UL2 model and studies
on ColBERT by Khattab and Zaharia (2020) have further pushed the boundaries of 
retrieval and generation synergies in RAG frameworks\cite{COLB}\cite{UL2}.

Despite the potential of RAG systems, their application within enterprise environments
remains under explored, particularly concerning the use of open-source 
solutions.Enterprises possess vast and diverse repositories of data, typically 
scattered across various internal systems and public websites.Proprietary LLMs, while 
powerful, come with high licensing costs and restrictive usage policies, limiting their
accessibility and adaptability for many organizations\cite{cost_analysis2024}. This creates a substantial 
barrier for enterprises, especially small to medium-sized ones, to leverage advanced 
Generative AI capabilities.

Evaluating the RAG framework involves several key metrics and methodologies to ensure 
its effectiveness and efficiency in real-world applications\cite{Eval}.There have been several 
metrics such as retrieval accuracy, response relevance, latency, and other metrics by 
others frameworks are crucial for assessing the performance of RAG systems\cite{RAGAS}\cite{deepeval}. Studies
have shown that the integration of advanced retrieval mechanisms with generative models
can significantly enhance the quality of responses in various Generative AI tasks. Additionally,
the adaptability of RAG systems to different data sets and contexts plays a vital role 
in their evaluation, highlighting the importance of domain-specific fine-tuning and 
optimization. This deepeval based evaluation framework helps in identifying the strengths and limitations of different RAG implementations, guiding the development of more robust 
and scalable solutions\cite{deepeval}.

Furthermore, our analysis of how TopK affects answer quality for different LLMs  reveals significant variations in performance. By adjusting the TopK parameter, which determines the number of top retrievals considered, we can observe changes in the accuracy and relevance of the generated answers. This analysis is crucial for understanding how to optimize retrieval settings for different models and use cases, providing insights into the fine-tuning required for optimal performance in specific enterprise contexts.

This research seeks to address the following things:

\begin{enumerate}    
    \item How do open-source LLMs and embeddings compare with each other and with proprietary alternatives in terms of accuracy and efficiency in RAG tasks using enterprise-specific data?
    \item What are the most important metrics to evaluate quality of RAG answers?
    \item What is the best combination of hyper parameters for RAG ?
\end{enumerate}

Addressing these questions is critical for providing enterprises with viable Generative AI solutions that are both effective and economically sustainable. This study will contribute to the broader understanding of the potential and limitations of open-source Generative AI tools in real-world business settings, offering a pathway for organizations to enhance their information retrieval and content generation capabilities.

\section{Methodology}
 This section outlines the methodology employed to evaluate the effectiveness of open-source large language models (LLMs) and embedding techniques in enhancing Retrieval-Augmented Generation (RAG) systems for enterprise-specific data. The methodology is structured into several key stages: data collection, model selection, system architecture, evaluation metrics, and experimental procedure.

\subsection{Data Collection}
\sloppy
Data collection is a crucial step in developing and evaluating Retrieval-Augmented 
Generation (RAG) systems, particularly when dealing with enterprise-specific datasets. 
In this study, data was scraped from the website \url{https://i-venture.org/}, utilizing a 
structured approach as mentioned below to ensure the accuracy and comprehensiveness of the collected data.

\subsubsection{Sitemap Extraction}
The initial step in the data collection process is extracting Uniform Resource Locator(URL) from the website’s sitemap. The process included:

\begin{enumerate}
    \item \textbf{URL Retrieval}: Accessing the sitemap located at \url{https://i-venture.org/sitemap.xml} to retrieve the Extensible Markup Language
(XML) content.
    \item \textbf{Parsing the Sitemap}: Using an XML parser to read and interpret the sitemap.
    \item \textbf{URL Extraction}: Extracting all \texttt{<loc>} tags, which contain the URLs of the web pages, and compiling these into a list. This list served as the foundation for the subsequent crawling process.
\end{enumerate}

\subsubsection{Web Crawling}
With the URLs extracted from the sitemap, each URL has to be crawled to extract textual content. This was performed using a breadth-first search approach with the following steps:

\begin{enumerate}
    \item \textbf{Queue Initialization}: A queue was initialized with the URLs obtained from the sitemap.
    \item \textbf{Directory Setup}: Directories were created to store raw text files and processed files, ensuring a well-organized data repository.
    \item \textbf{Content Extraction}: For each URL in the queue, the web page content was fetched using HTTP requests. The HTML content was parsed using an HTML parser to extract the main textual content while ignoring the HTML tags.
    \item \textbf{File Storage}: The extracted text was cleaned (e.g., by removing extra white spaces ,new line characters and special characters ) and saved into text files named based on the URL structure. This involved replacing slashes and other non-filename characters with underscores to ensure valid file names.
\end{enumerate}

This ensures that the dataset from \url{https://i-venture.org} was comprehensive, clean, and ready to be used for RAG task.

\subsection{Text Splitting}

The next step is splitting the data into chunks. This is essential for ensuring that the text is appropriately segmented so that only the most relevant chunks gets passed to the LLM for RAG. The process was carried out using the \texttt{langchain} library, specifically leveraging the \texttt{DirectoryLoader} and \texttt{NLTKTextSplitter} tools\cite{langchain}.

\begin{enumerate}
    \item \textbf{DirectoryLoader}: This tool was directed to the directory containing all the text files to load all the text.
    \item \textbf{NLTKTextSplitter}: This tool was used to divide the text into smaller chunks based on NLTK tokens without trying to break paragraphs sentences and words. Smaller text chunks improve the accuracy of the retrieval component by allowing it to match queries more precisely with relevant sections of text\cite{nltk}. For NLTKSplitter chunk size need not be specified.
    However TopK value of 5 has been used for this evaluation while using NLTKSplitter.
    \item \textbf{RecursiveCharacterTextSplitter}:  This can also be used which splits the text into chunks as small as possible while preserving the structure of all paragraphs (and then sentences, and then words) together , as those would generically seem to be the strongest semantically related pieces of text. It also uses chunk overlap ensuring information is available between neighbouring chunks\cite{recr}.
In this study for the evaluation the RecursiveCharacterTextSplitter with token limit of 1024 and chunk overlap of 102 tokens has been used.
\end{enumerate}

\subsection{Embedding Generation}

The next step involves generating embeddings for the text chunks created by text splitting. Embeddings are crucial for converting text into numerical representations that can be efficiently processed by llms\cite{MTEB}. This study utilized embeddings from Hugging Face, a popular platform providing pre-trained models for various Generative AI tasks\cite{huggingface}.

\begin{enumerate}
    \item \textbf{Embedding Storage}: A local file store was set up to cache the embeddings, ensuring efficient access and retrieval during RAG.
    \item \textbf{Embedding Model}: In this work the required embedding model has been loaded from the Hugging Face\cite{huggingface}.
    \textbf{BAAI/bge-large-en-v1.5} has been selected as the embedding model owing to it's good performance in semantic search\cite{bge_embedding} . Besides it also supports ReRanking of retrieved texts \cite{BAAI}.
\end{enumerate}

\subsection{Vector Database Creation}

 FAISS (Facebook AI Similarity Search) has been used to create a vector database to store embeddings \cite{faiss}. The text chunks created before were converted into embeddings and stored in this FAISS vector database. This creates a structured repository that supports fast retrieval based on semantic similarity.

\subsection{LLM Integration}
\sloppy
This step involves incorporating LLMs to enhance the generative component of the Retrieval-Augmented Generation (RAG) system. For this study, we utilized open-source LLMs provided by Perplexity, integrating them into the \texttt{langchain} framework through a custom wrapper function (git link to wrapper code) . This integration was done by adapting the approach detailed in this github repository \cite{wrapper}\cite{perp}.

\subsubsection{Perplexity API}

Perplexity offers API to a wide range of powerful open-source LLMs that can generate human-like text, making them ideal for tasks such as text generation, summarization, and question answering \cite{perplex}. These models were chosen for their accessibility and flexibility, which are essential for enterprise applications where commercial/proprietary models might be prohibitively expensive\cite{cost}.

\subsubsection{Benefits of Using Perplexity API}

\begin{itemize}
    \item \textbf{Cost-Effectiveness}: As open-source models, Perplexity provide a cost-effective alternative to proprietary solutions. For example, GPT-3.5 costs around 2 USD per million tokens on average, Perplexity LLMs takes only 0.6 USD per million tokens which is less than one-third of the price\cite{cost}\cite{perpcost}.
    \item \textbf{Accessibility}: Perplexity provides API access to these open source LLMs without having to invest in local GPU capacity\cite{neur}\cite{scal}.
\end{itemize}

\subsection{Retrieval-Augmented Generation (RAG)}

The next stage involves implementing the Retrieval-Augmented Generation (RAG) framework. This framework combines powerful retrieval techniques with generative models to produce accurate and contextually relevant outputs. The RAG system implemented in this work utilizes a hybrid retriever approach, combining a custom Best Match 25(BM25) retriever with a FAISS-based vector retriever, followed by a question-answering (QA) module that generates responses based on the retrieved information \cite{BM25}.

\subsubsection{Hybrid Retriever Setup}

To enhance retrieval accuracy, a hybrid retriever approach was adopted. This involved using both BM25 and FAISS retrievers with same weightage to both retrievers. 

\begin{itemize}
    \item \textbf{BM25 Retriever}: The BM25 algorithm is a well-known probabilistic information retrieval model that ranks documents based on their relevance to a query. It was configured to retrieve the top 5 documents most relevant to each query.
    \item \textbf{FAISS Retriever}: The FAISS retriever, leveraging the vector embeddings generated in the previous steps, was also set to retrieve the top k documents. FAISS excels in handling large-scale similarity searches efficiently.
    \item \textbf{Ensemble Retriever}: An ensemble retriever was created to combine the results from both BM25 and FAISS retrievers. Each retriever was assigned equal weight, ensuring a balanced contribution from both methods. This type of
hybrid retriever has demonstrated better performance compared to a vector retriever alone \cite{cos}. 
\end{itemize}

\subsubsection{RetrievalQA}

After setting up the hybrid retriever, the next step was to implement the QA module using the \texttt{RetrievalQA} chain from the \texttt{langchain} library. This module is designed to generate answers to queries by leveraging the retrieved documents and providing source references for the generated answers.

\begin{itemize}
    \item \textbf{Retriever Integration}: The hybrid retriever was used in conjunction with the QA module to ensure that the generated answers were based on the most relevant and contextually appropriate documents.
    \item \textbf{Callback Handling}: A callback handler was integrated to facilitate real-time monitoring and debugging of the QA process.
    %%\item \textbf{Chain Configuration}: The QA chain was configured with a custom prompt and set to return the source documents along with the generated answers. This setup ensures transparency and traceability of the information provided by the system.
\end{itemize}

\subsection{Evaluation}

The evaluation of the Retrieval-Augmented Generation (RAG) system involves multiple metrics to assess its performance comprehensively. This includes measuring the quality of the generated responses, the efficiency of the retrieval process, and the contextual relevance of the answers. The following methods were employed:

\subsubsection{ROUGE Scores}

ROUGE (Recall-Orieated Understudy for Gisting Evaluation) is a set of metrics commonly used for evaluating the quality of text in tasks such as summarization and machine translation. It compares the overlap of n-grams between the generated text and the reference text.

\subsubsection{DeepEval Metrics}

To further assess the contextual quality of the responses, the DeepEval framework was used. This framework provides metrics for evaluating the precision, recall, and relevancy of the generated text in relation to the expected outputs and the context provided by the retrieved documents\cite{deepeval}.

\begin{itemize}
    \item \textbf{Contextual Precision}: Measures how many of the retrieved documents contain information that is relevant to the generated response.
    \item \textbf{Contextual Recall}: Measures the proportion of relevant information in the retrieved documents that is used in the generated response.
    \item \textbf{Contextual Relevancy}: Assesses how relevant the generated response is to the query and the context provided by the retrieved documents.
\end{itemize}

The evaluation process involved creating test cases where the input query, the actual output from the system, the expected output (generated using GPT-4), and the retrieval context were used to measure these metrics. Research has shown that GPT-4 agrees with human labelers around 80\% of the time, making it a reliable and scalable option for evaluating natural language output \cite{llmllm}.

The evaluation combines both qualitative and quantitative measures to provide a comprehensive assessment of the RAG system's performance. By using ROUGE scores, inference time, and DeepEval metrics, this study ensures that the system is evaluated for its accuracy, efficiency, and contextual relevance. %%This multifaceted evaluation approach helps in identifying strengths and areas for improvement, contributing to the development of more effective and reliable open-source RAG system alternatives for enterprise applications.

\section{Results}
\label{sec:Result}
When evaluating a dataset, categorizing the evaluation set into different segments helps in understanding the performance of models under various conditions. Here, the categories are based on the density of reasoning and factual information. Each category is defined as follows:
\begin{itemize}
    \item \textbf{Reason Dense:} Reason dense segments consist of reasoning-based questions where the reasoning-related information appears repeatedly throughout the dataset. This involves complex reasoning tasks with repeated reasoning patterns or information.
    
    \item \textbf{Reason Sparse:} Reason sparse segments include reasoning-based questions where the reasoning-related information appears infrequently. This involves simpler reasoning tasks with limited instances of reasoning information.
    
    \item \textbf{Factual Dense:} Factual dense segments consist of factual questions where detailed factual information is repeated many times in the dataset. This involves numerous repeated facts requiring detailed knowledge.
    
    \item \textbf{Factual Sparse:} Factual sparse segments include factual questions where the factual information appears infrequently. This involves general knowledge with minimal repeated factual details.
\end{itemize}
\subsection{Evaluating Answer Quality}
For this evaluation, we selected three questions from each category of the above mentioned 4 categories and RecursiveCharacterTextSplitter has been used. The performance of the open sources LLMs used in RAG framework has been evaluated by calculating various metrics viz:
\subsubsection{Cosine similarity} The cosine similarity metric measures the cosine of the angle between two non-zero vectors, providing a similarity score that ranges from -1 to 1. A higher score indicates greater similarity between the query and the content. In this work the cosine similarity between score between the query and the retrieved context for various top-k values has been calculated,where top-k denote the number of top ranking documents extracted during retrieval.
%%multiple scores cosine similarity scores between the query and the extracted content for various top-k values. The cosine similarity metric measures the cosine of the angle between two non-zero vectors, providing a similarity score that ranges from -1 to 1. A higher score indicates greater similarity between the query and the content. In our experiments, we computed the cosine similarity scores for different top-k values, which denote the number of top-ranking documents considered during retrieval.

\begin{figure*}[ht]
    \centering
    \begin{minipage}{0.48\textwidth} % Adjusted width of the minipage
        \centering
        \begin{subfigure}{0.48\textwidth} % Adjusted width of subfigures
            \includegraphics[width=\linewidth]{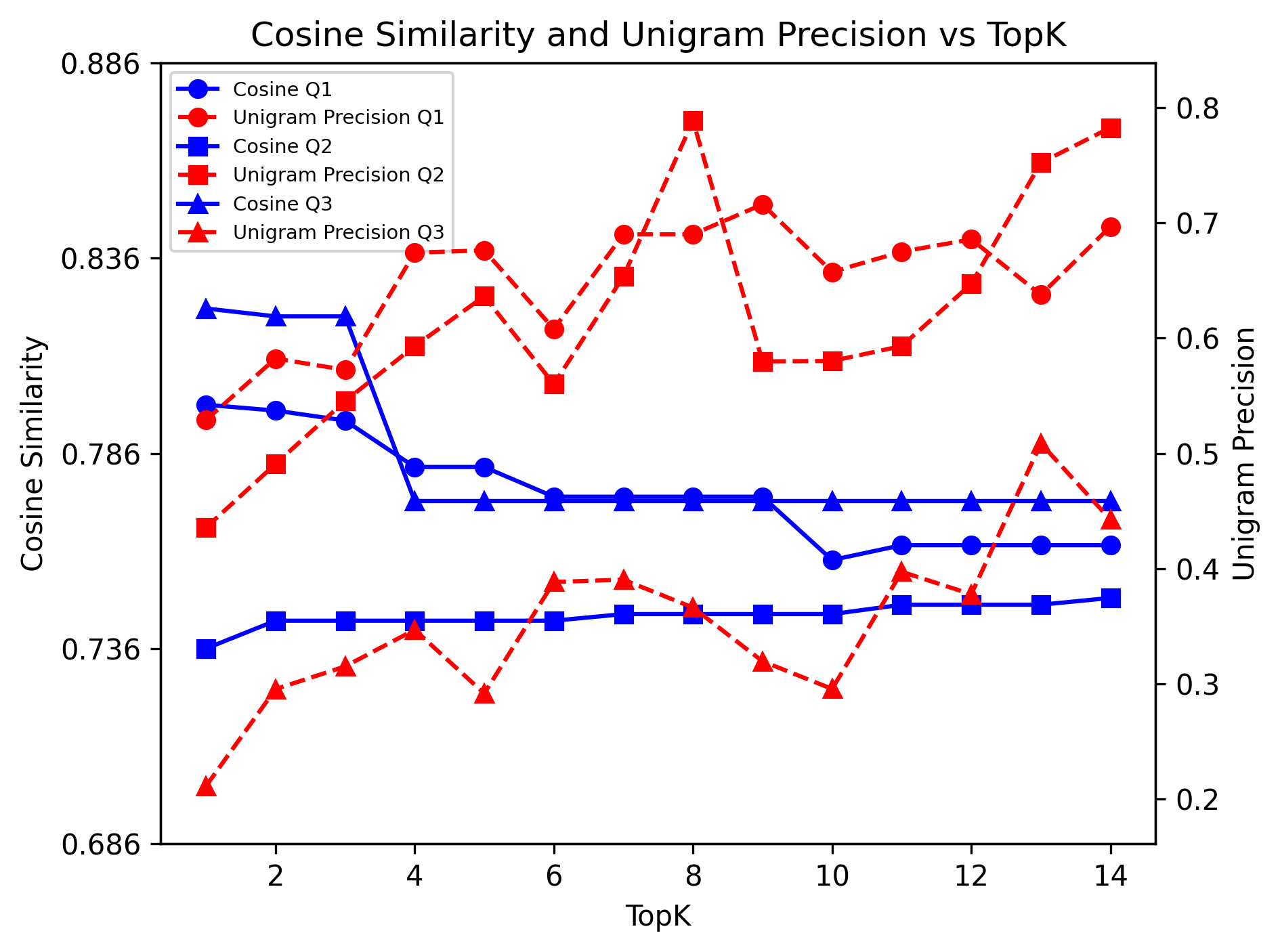}
            \caption{Mistral: Reason Dense}
            \label{fig:mistral_reson_dense_reason_sparse}
        \end{subfigure}
        \hfill
        \begin{subfigure}{0.48\textwidth}
            \includegraphics[width=\linewidth]{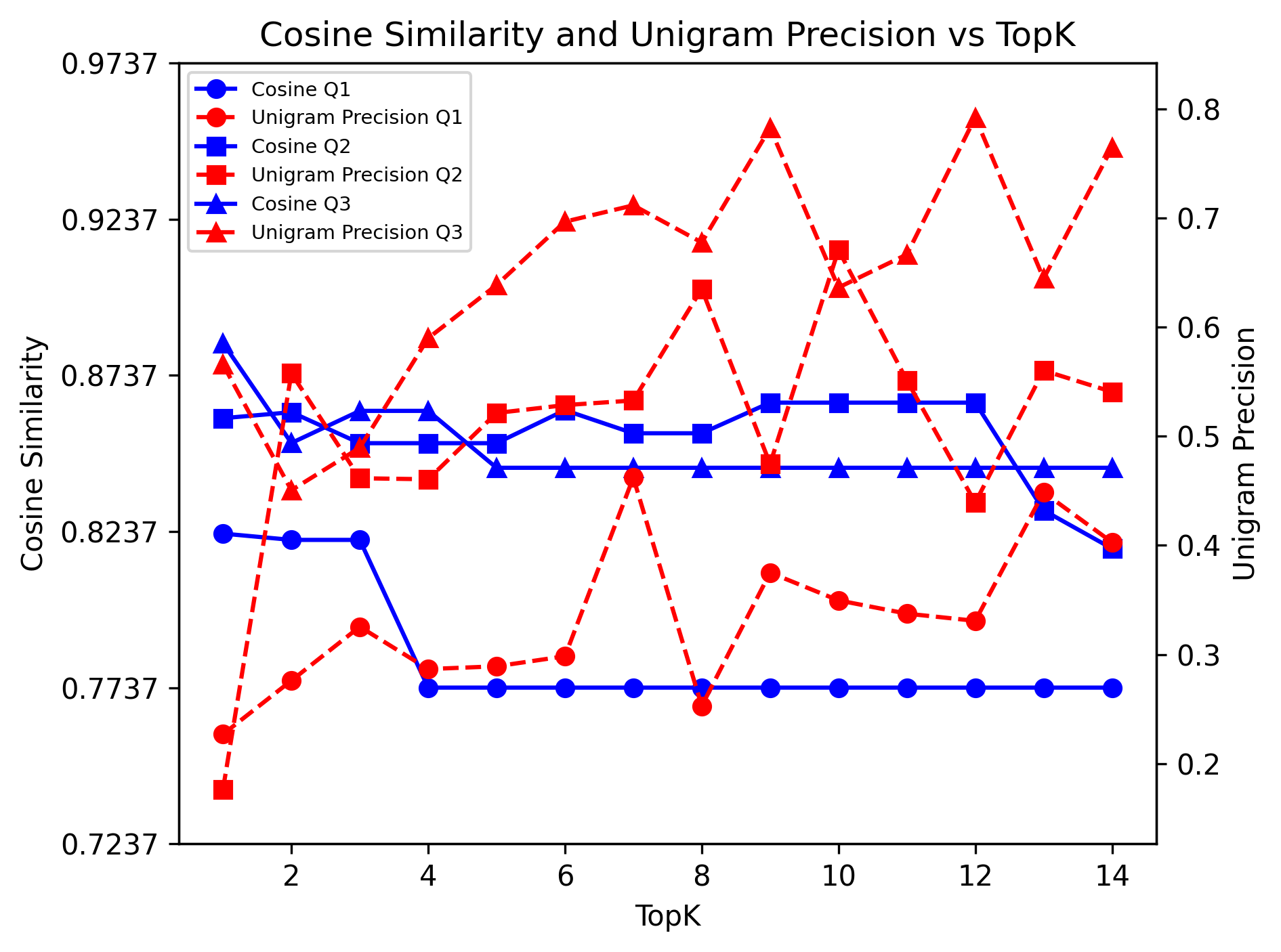}
            \caption{Mistral: Reason Sparse}
            \label{fig:mistral_factual_dense_factual_sparse}
        \end{subfigure}
        \vspace{0.01cm} % Adjust vertical space between rows
        \begin{subfigure}{0.48\textwidth}
            \includegraphics[width=\linewidth]{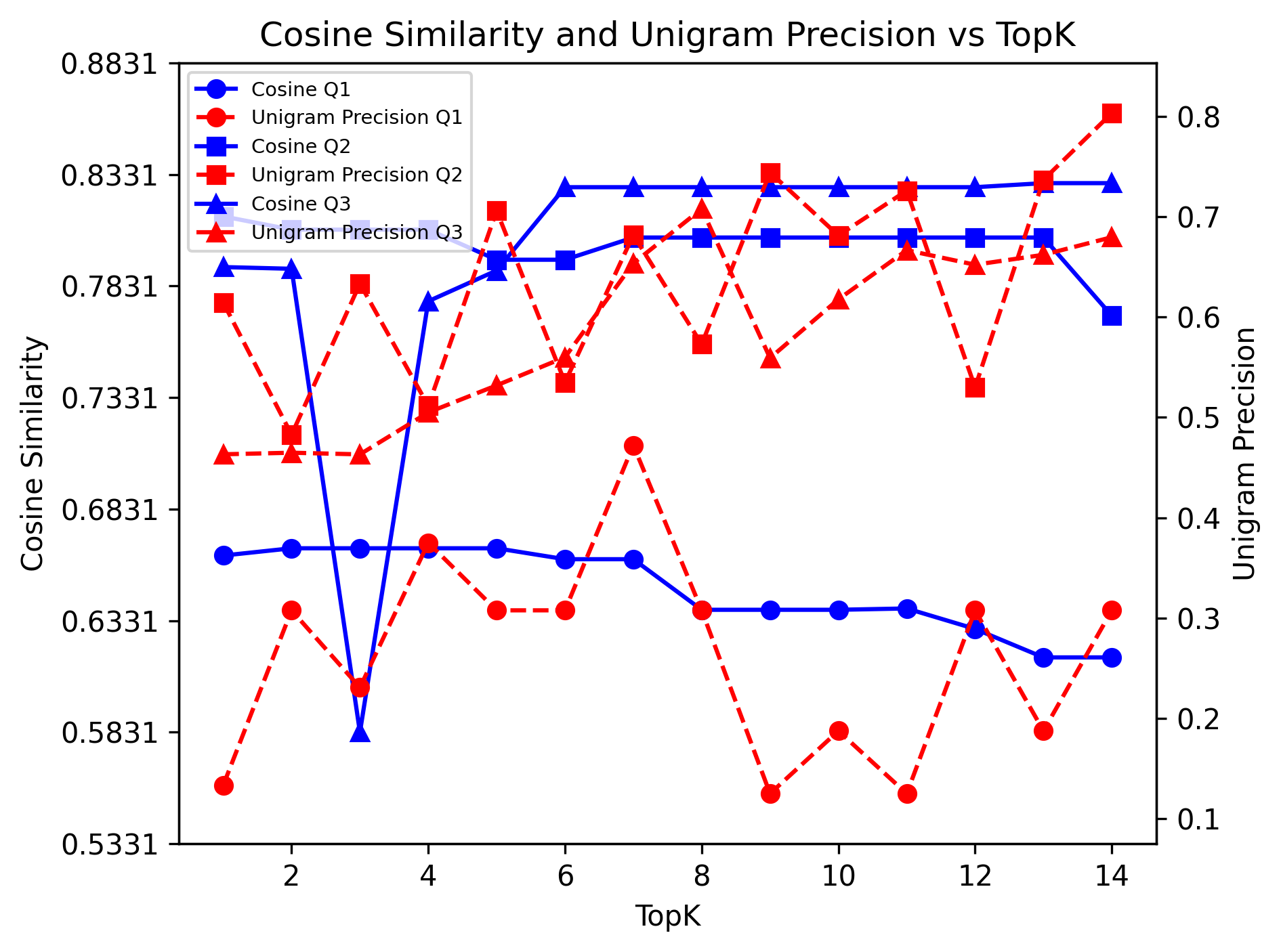}
            \caption{Mistral: Factual Dense}
            \label{fig:mistral_reson_dense_reason_sparse_trendlines}
        \end{subfigure}
        \hfill
        \begin{subfigure}{0.48\textwidth}
            \includegraphics[width=\linewidth]{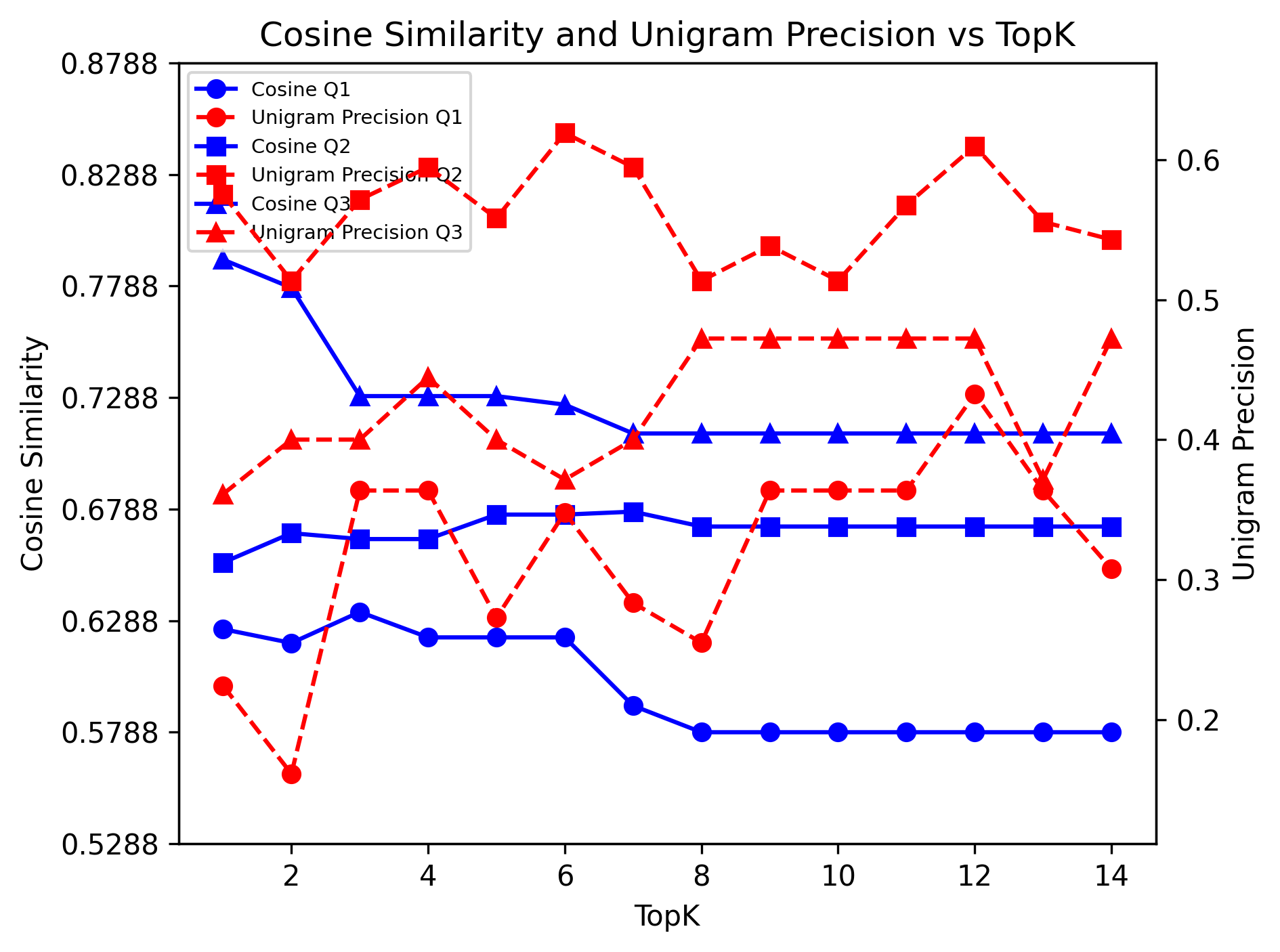}
            \caption{Mistral: Factual Sparse}
            \label{fig:mistral_factual_dense_factual_sparse_trendlines}
        \end{subfigure}
    \end{minipage}
    \hfill
    \begin{minipage}{0.48\textwidth} % Adjusted width of the minipage
        \centering
        \begin{subfigure}{0.48\textwidth} % Adjusted width of subfigures
            \includegraphics[width=\linewidth]{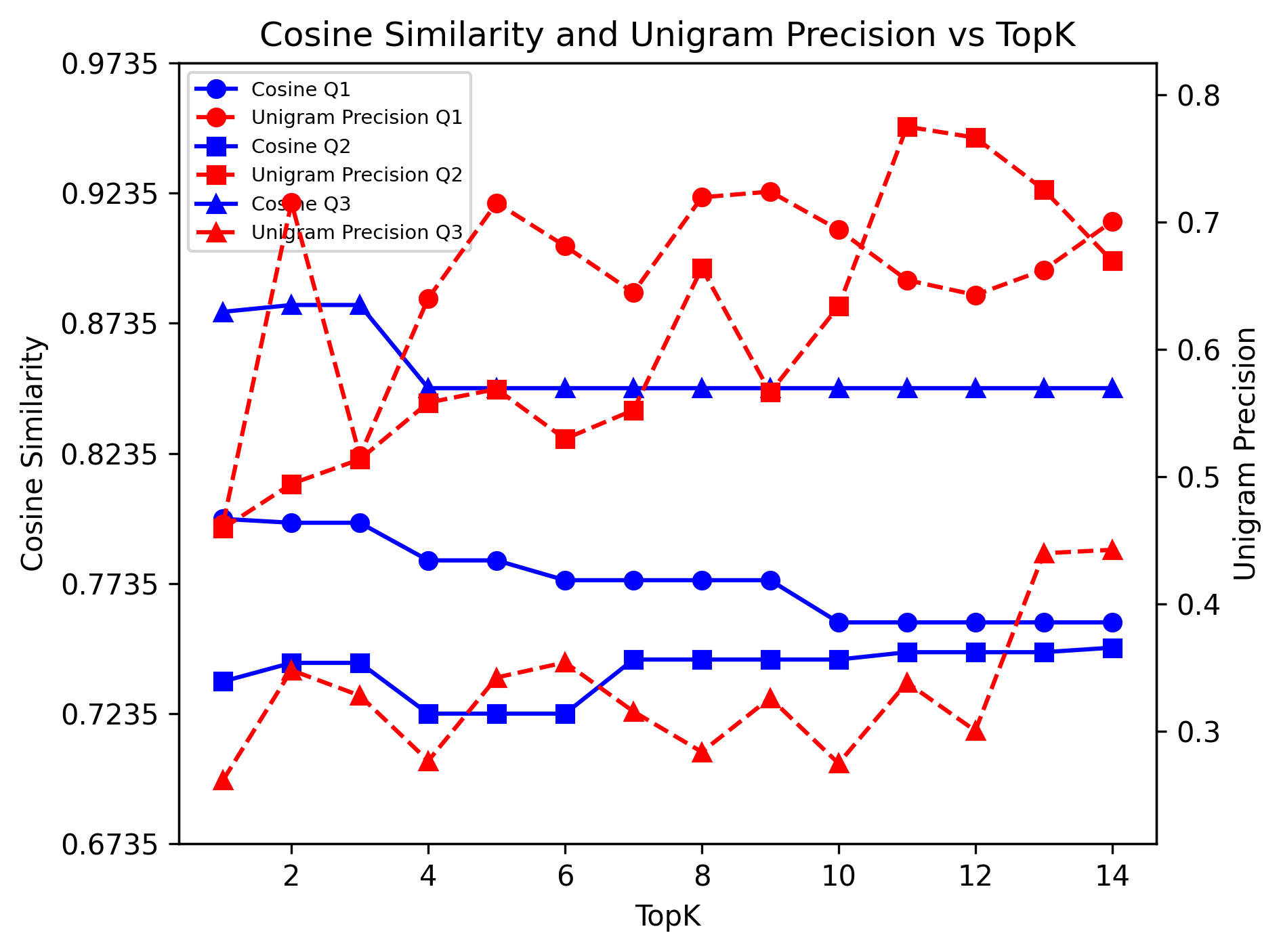}
            \caption{LLama3: Reason Dense}
            \label{fig:llama_reson_dense_reason_sparse}
        \end{subfigure}
        \hfill
        \begin{subfigure}{0.48\textwidth}
            \includegraphics[width=\linewidth]{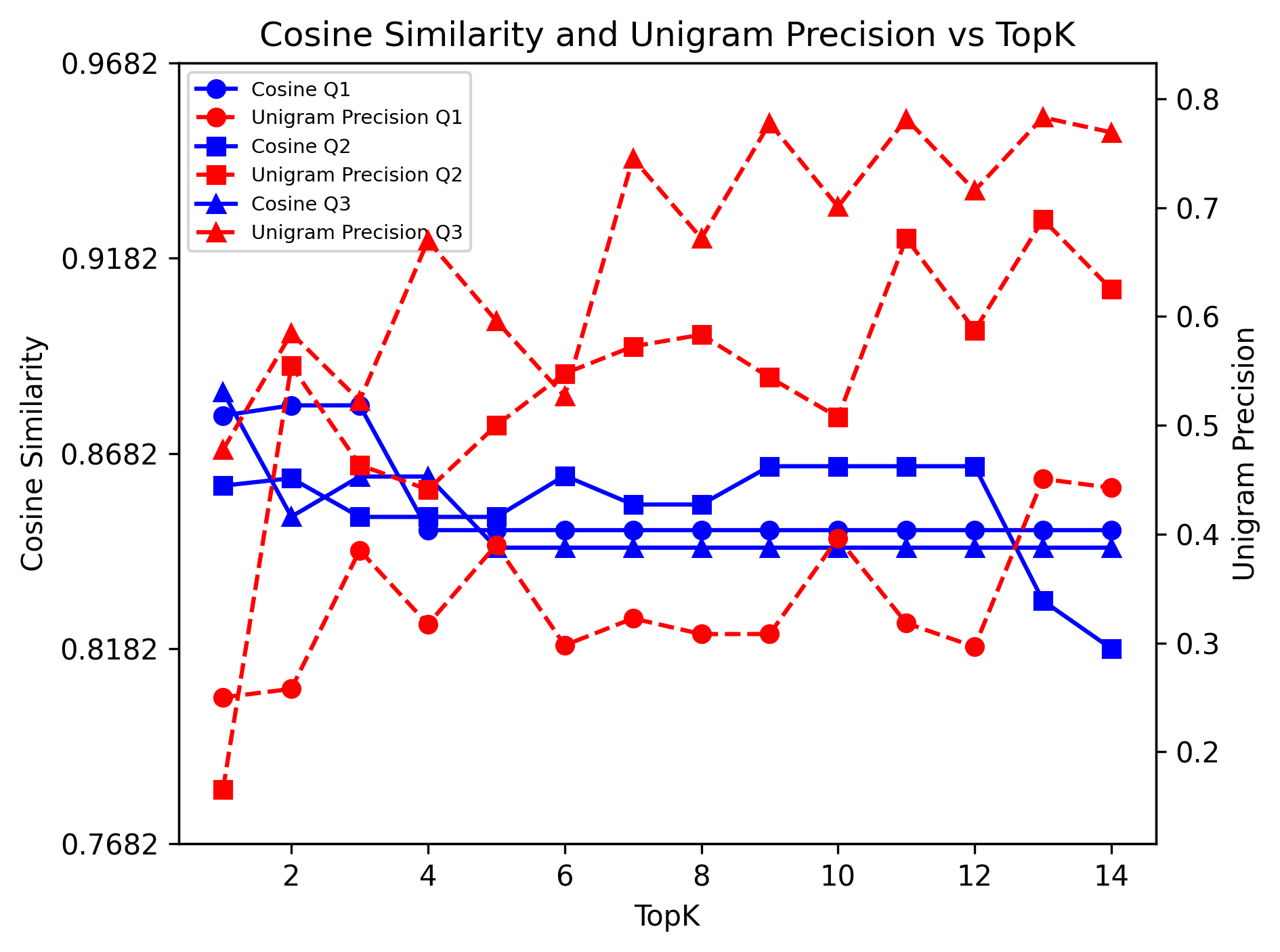}
            \caption{LLama3: Reason Sparse}
            \label{fig:llama_factual_dense_factual_sparse}
        \end{subfigure}
        \vspace{0.1cm} % Adjust vertical space between rows
        \begin{subfigure}{0.48\textwidth}
            \includegraphics[width=\linewidth]{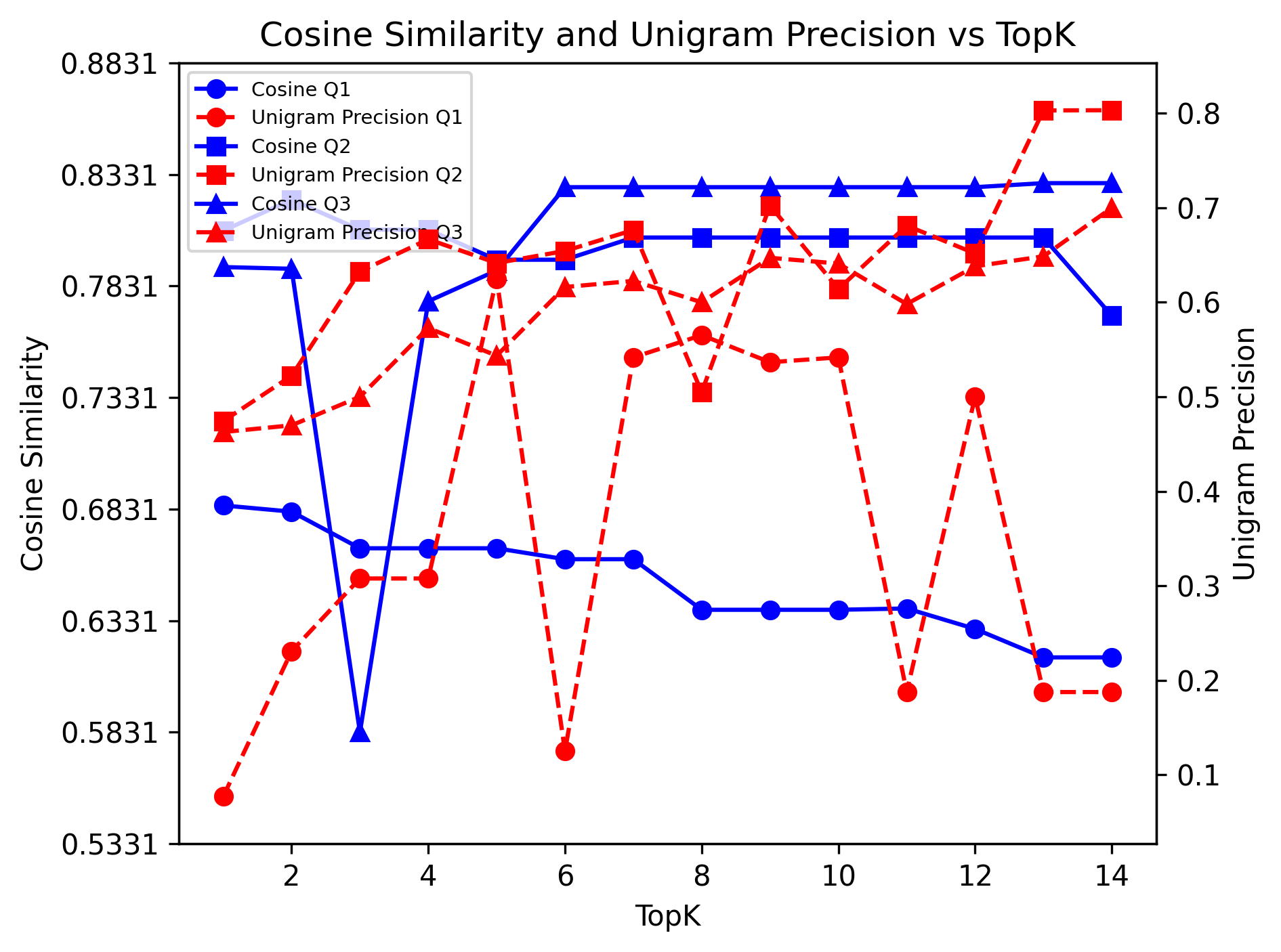}
            \caption{LLama3: Factual Dense}
            \label{fig:llama_reson_dense_reason_sparse_trendlines}
        \end{subfigure}
        \hfill
        \begin{subfigure}{0.48\textwidth}
            \includegraphics[width=\linewidth]{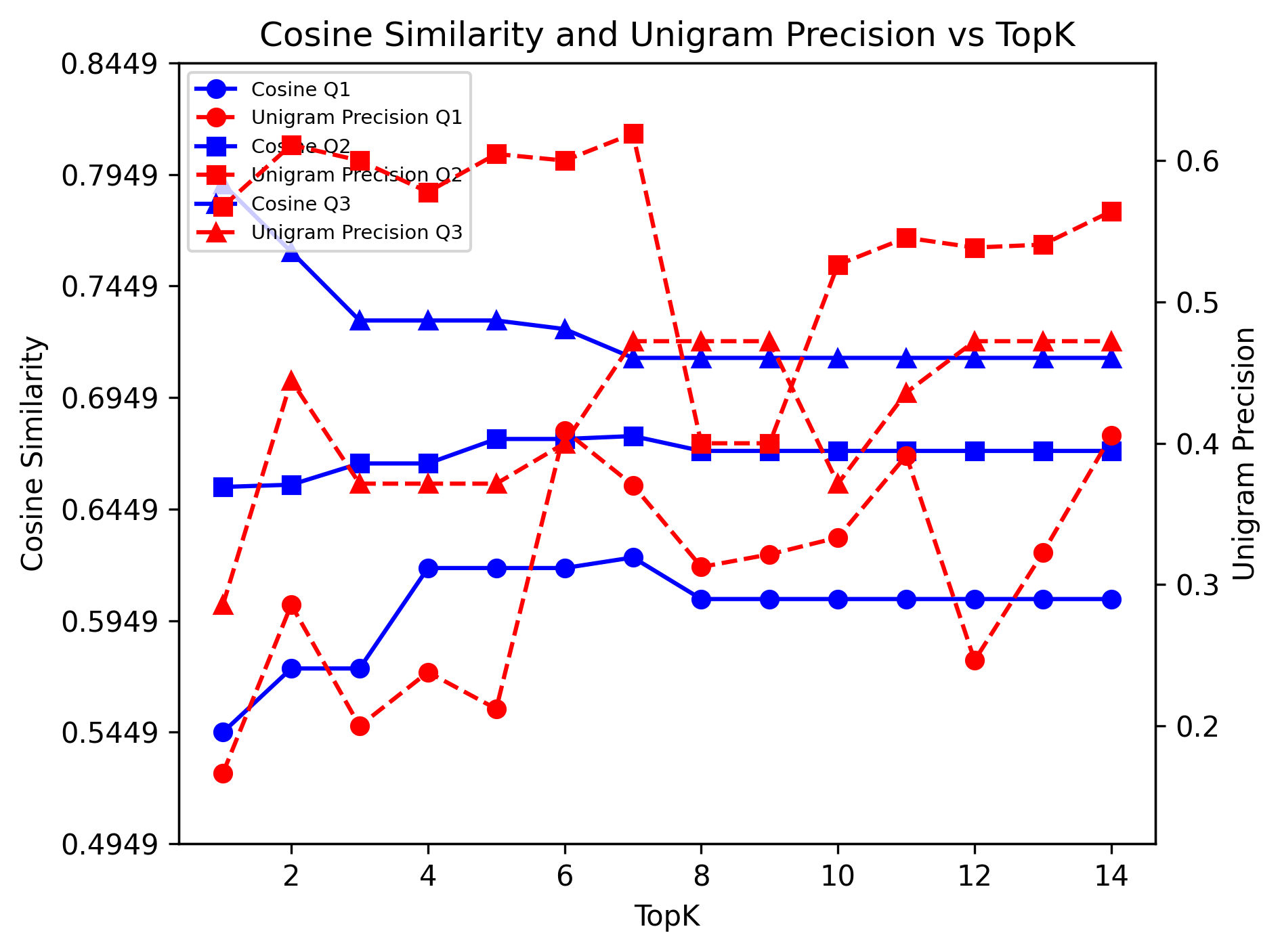}
            \caption{LLama3: Factual Sparse}
            \label{fig:llama_factual_dense_factual_sparse_trendlines}
        \end{subfigure}
    \end{minipage}
    \caption{Analysis of Cosine Similarity and Unigram Precision vs TopK for Mistral8x7B LLM model (left) and LLama3-8B LLM model (right)}
    \label{fig:combined_plots_2}
\end{figure*}

Based on the analysis presented in Figure \ref{fig:combined_plots_2}, it is observed that as the value of top-k increases, the cosine similarity scores exhibit a trend of initial increase, followed by a plateau as top-k continues to rise. This suggests a diminishing return on similarity with higher top-k values. 
\subsubsection{Unigram Precision} : Unigram precision measures the fraction of the words  in the LLM generated answer that are also present in the reference/context.
Conversely, unigram precision shows a more erratic pattern but generally demonstrates an upward trend initially, reaching a peak before declining. This indicates that while increasing top-k can enhance unigram precision up to a certain point, further increments may lead to reduced precision. Additionally, it is noted that as cosine similarity starts to decrease, there are instances where unigram precision also dips across all four types of questions, further highlighting the interplay between similarity and precision metrics.

The Precision vs top-k graphs did not exhibit consistent variations across the Llama3 and Mistral models, as shown in Figure \ref{fig:combined_plots_2} . For Reason Dense data, the Llama3 model displayed an initial peak followed by a decrease for one question, whereas Mistral demonstrated a steady increase. In the Reason Sparse category, Llama3 showed a gradual increase, while Mistral peaked and then declined for two questions. For Factual Dense data, both models generally showed a steady increase, except for one question where Mistral initially decreased sharply before rising again. In the Factual Sparse category, the initial peak for Llama3 was not achieved again, whereas Mistral remained relatively constant. These observations highlight the different behaviors of the two models across various data types and question contexts.

\begin{figure*}[ht]
    \centering
    \begin{minipage}{0.48\textwidth} % Slightly reduced width of the minipage
        \centering
        \begin{subfigure}{0.48\textwidth} % Adjusted width of subfigures
            \includegraphics[width=\linewidth]{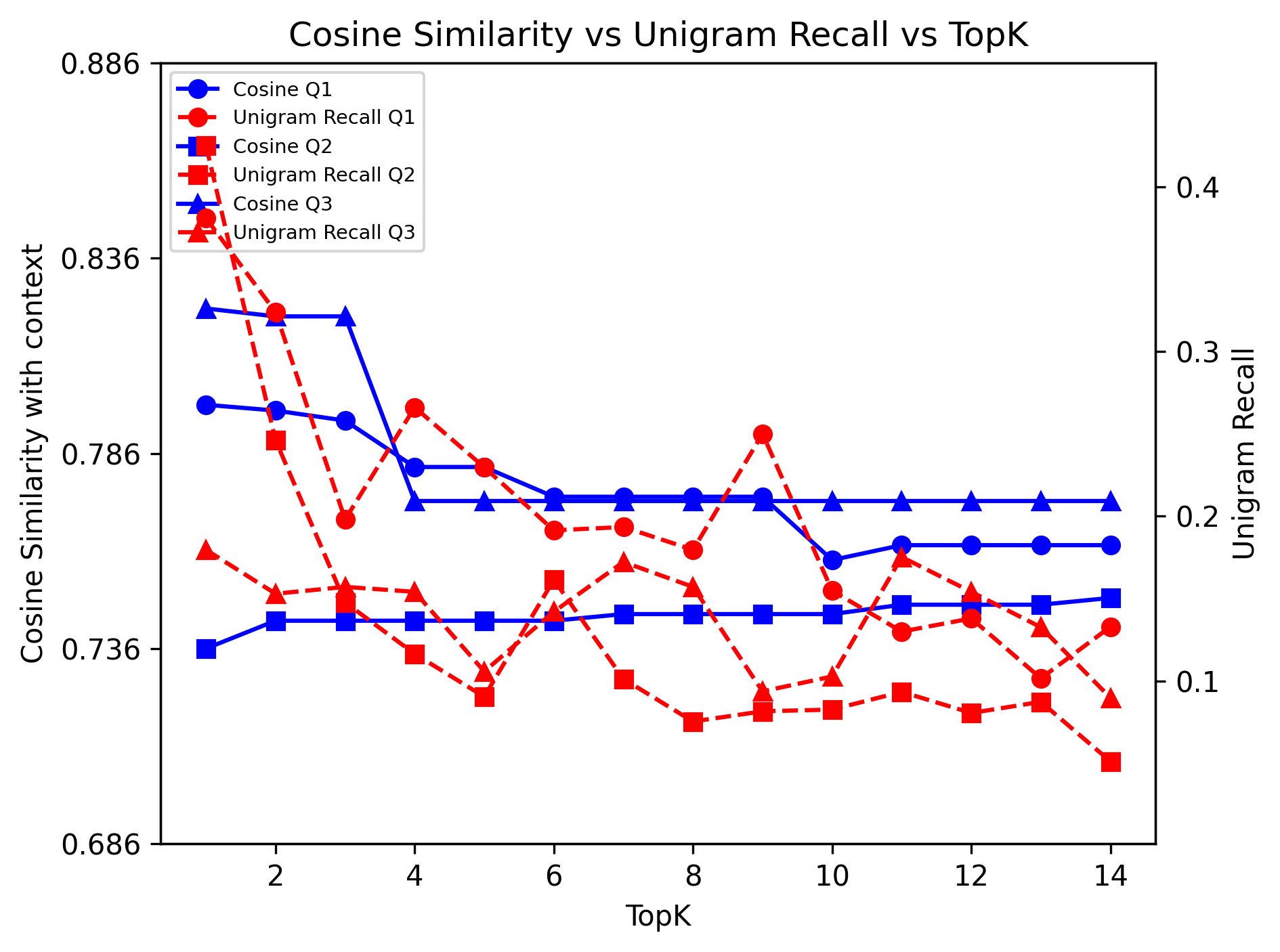}
            \caption{Mistral: Reason Dense}
            \label{fig:mistral_reson_dense}
        \end{subfigure}
        \hfill
        \begin{subfigure}{0.48\textwidth}
            \includegraphics[width=\linewidth]{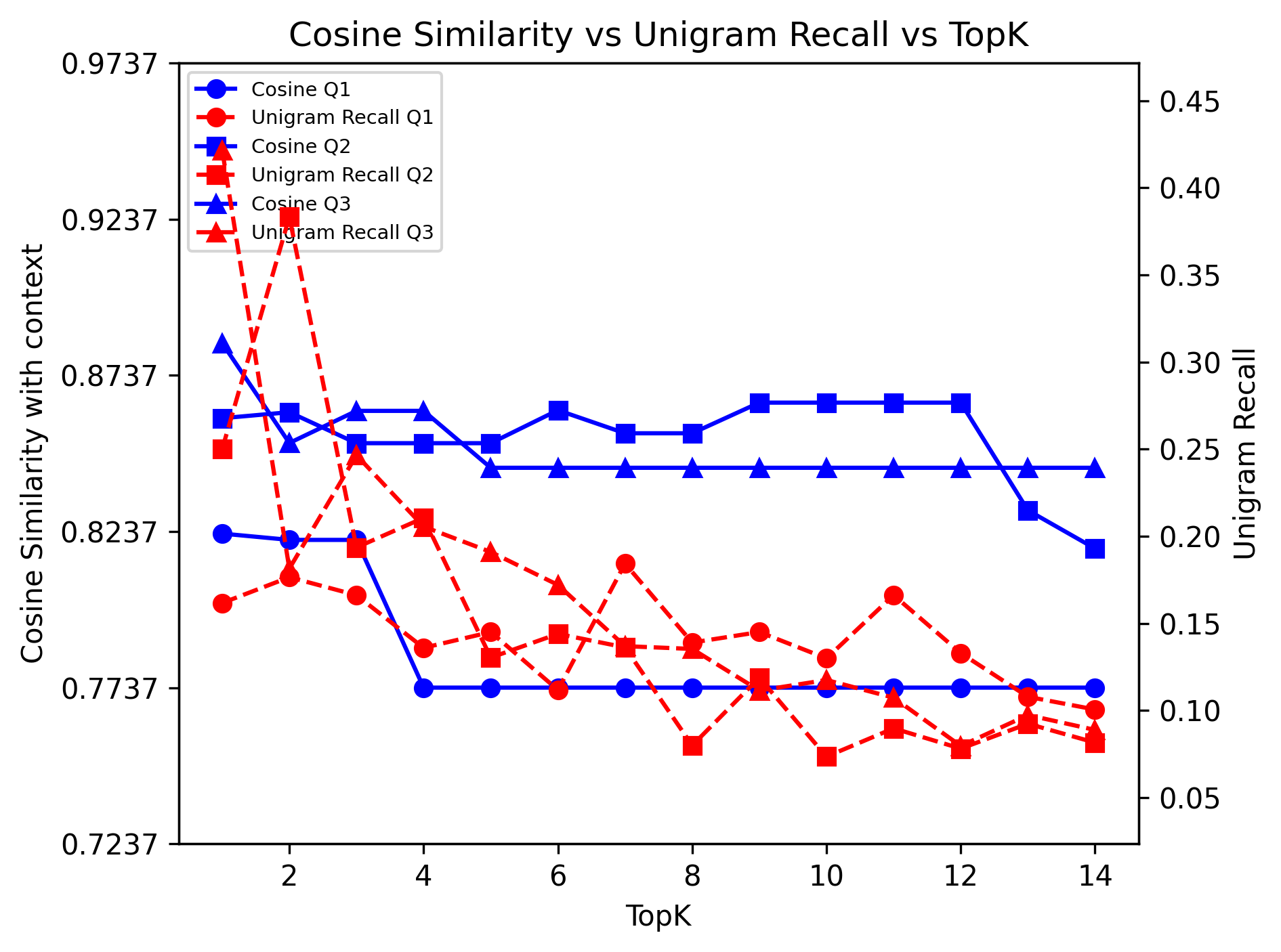}
            \caption{Mistral: Reason Sparse}
            \label{fig:mistral_reason_sparse}
        \end{subfigure}
        \vspace{0.1cm} % Adjust vertical space between rows
        \begin{subfigure}{0.48\textwidth}
            \includegraphics[width=\linewidth]{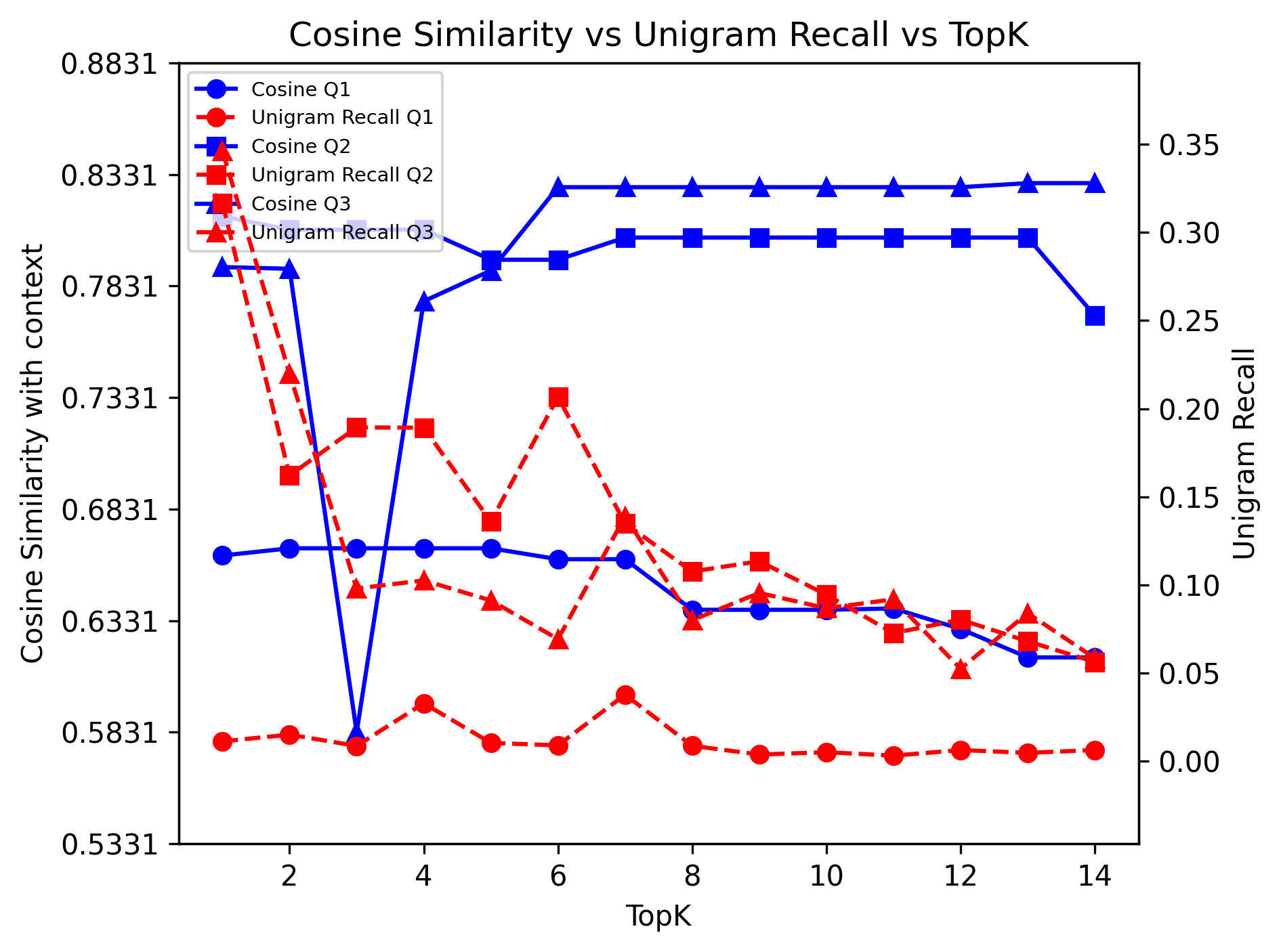}
            \caption{Mistral: Factual Dense}
            \label{fig:mistral_factual_dense}
        \end{subfigure}
        \hfill
        \begin{subfigure}{0.48\textwidth}
            \includegraphics[width=\linewidth]{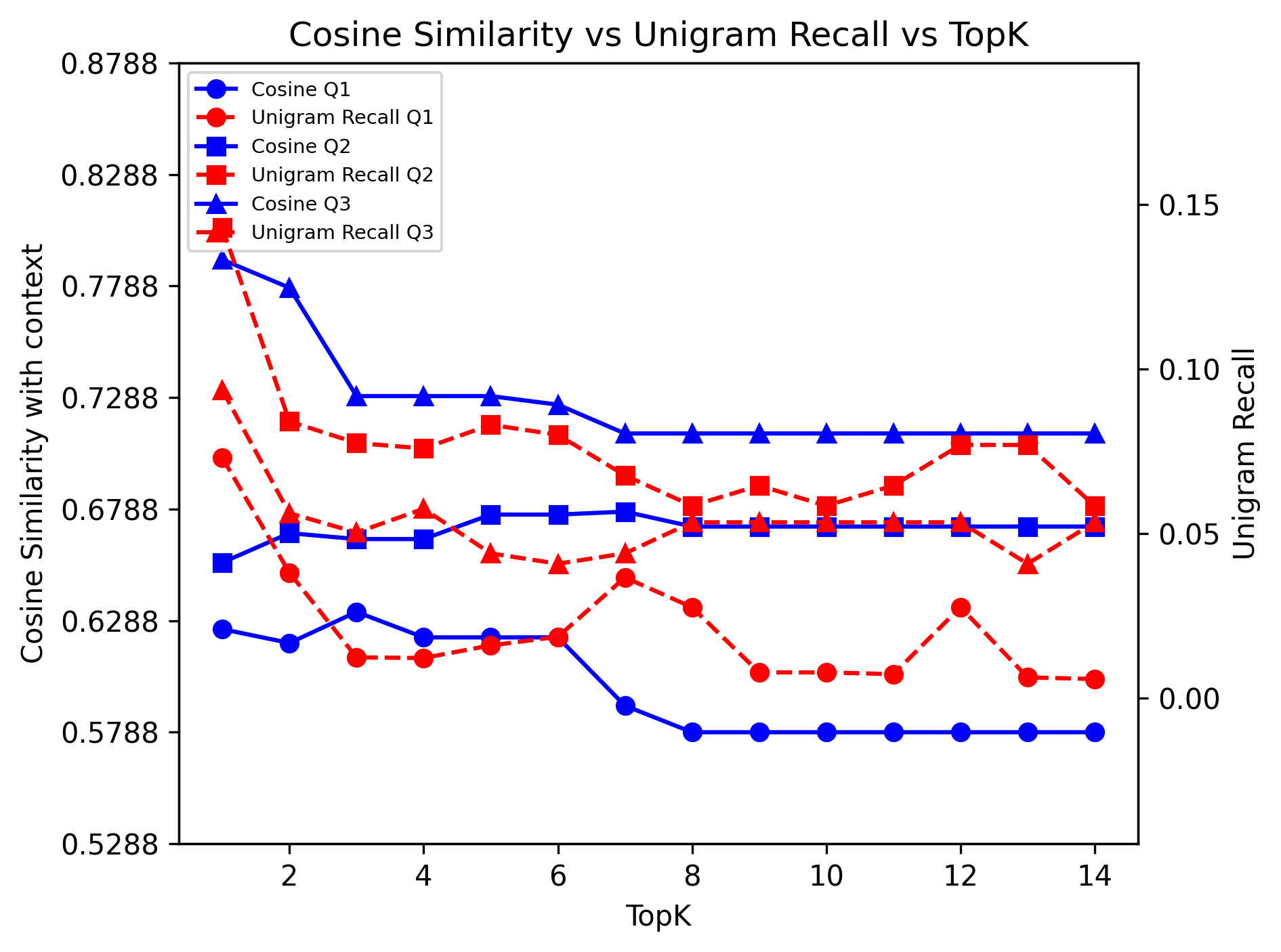}
            \caption{Mistral: Factual Sparse}
            \label{fig:mistral_factual_sparse}
        \end{subfigure}
    \end{minipage}
    \hfill
    \begin{minipage}{0.48\textwidth} % Slightly reduced width of the minipage
        \centering
        \begin{subfigure}{0.48\textwidth} % Adjusted width of subfigures
            \includegraphics[width=\linewidth]{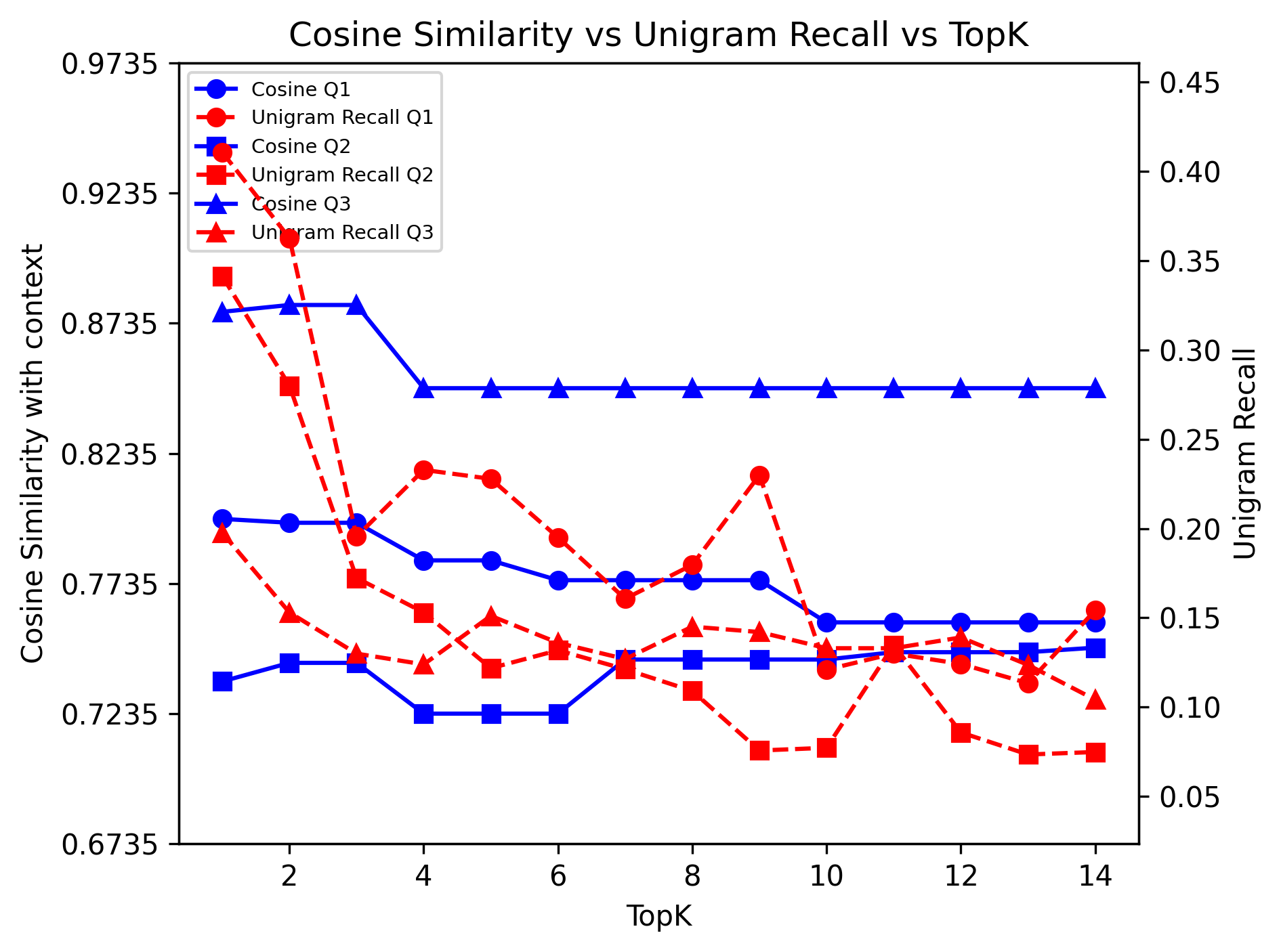}
            \caption{Llama3: Reason Dense}
            \label{fig:llama_reson_dense}
        \end{subfigure}
        \hfill
        \begin{subfigure}{0.48\textwidth}
            \includegraphics[width=\linewidth]{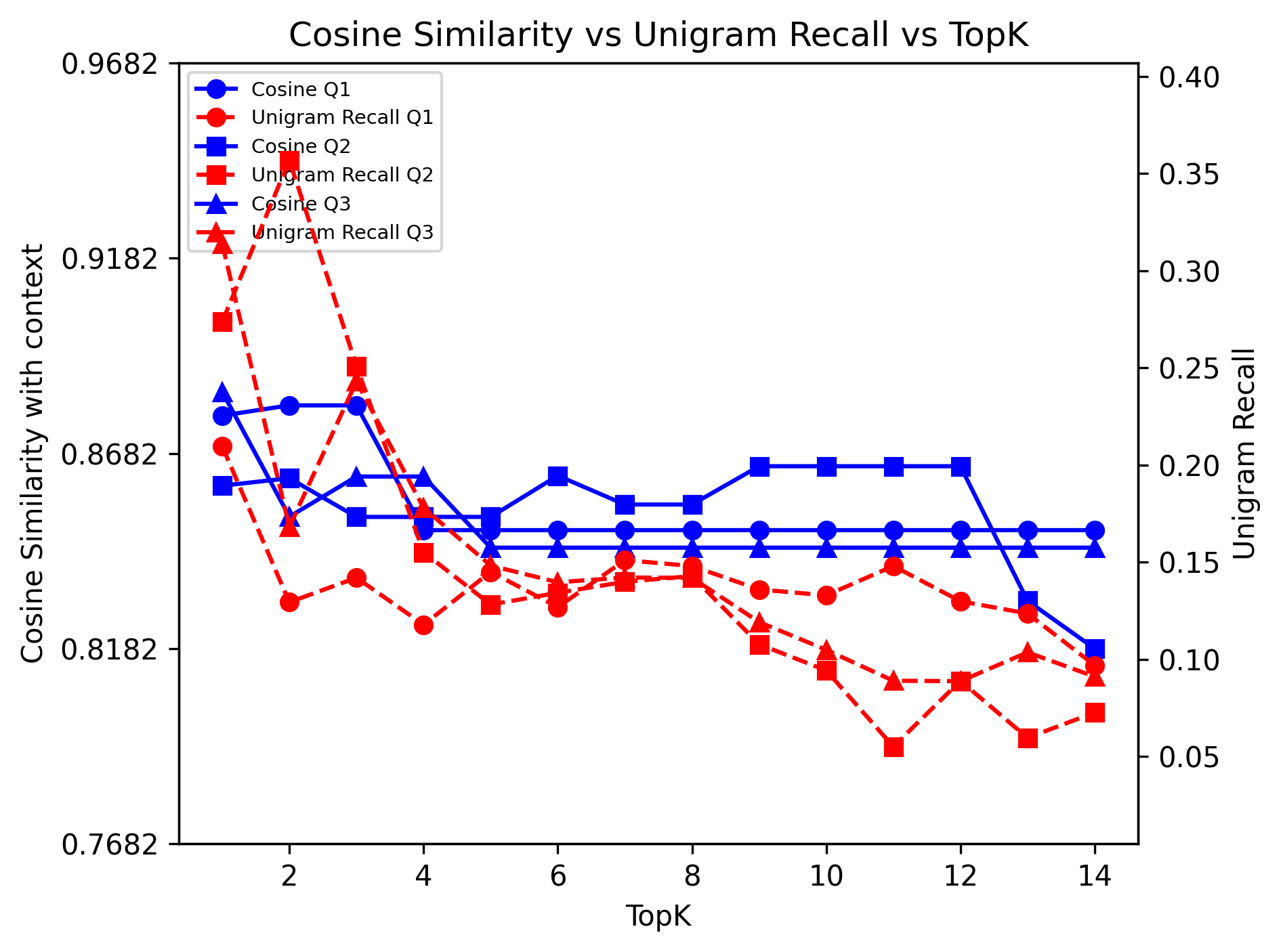}
            \caption{Llama3: Reason Sparse}
            \label{fig:llama_reason_sparse}
        \end{subfigure}
        \vspace{0.2cm} % Adjust vertical space between rows
        \begin{subfigure}{0.48\textwidth}
            \includegraphics[width=\linewidth]{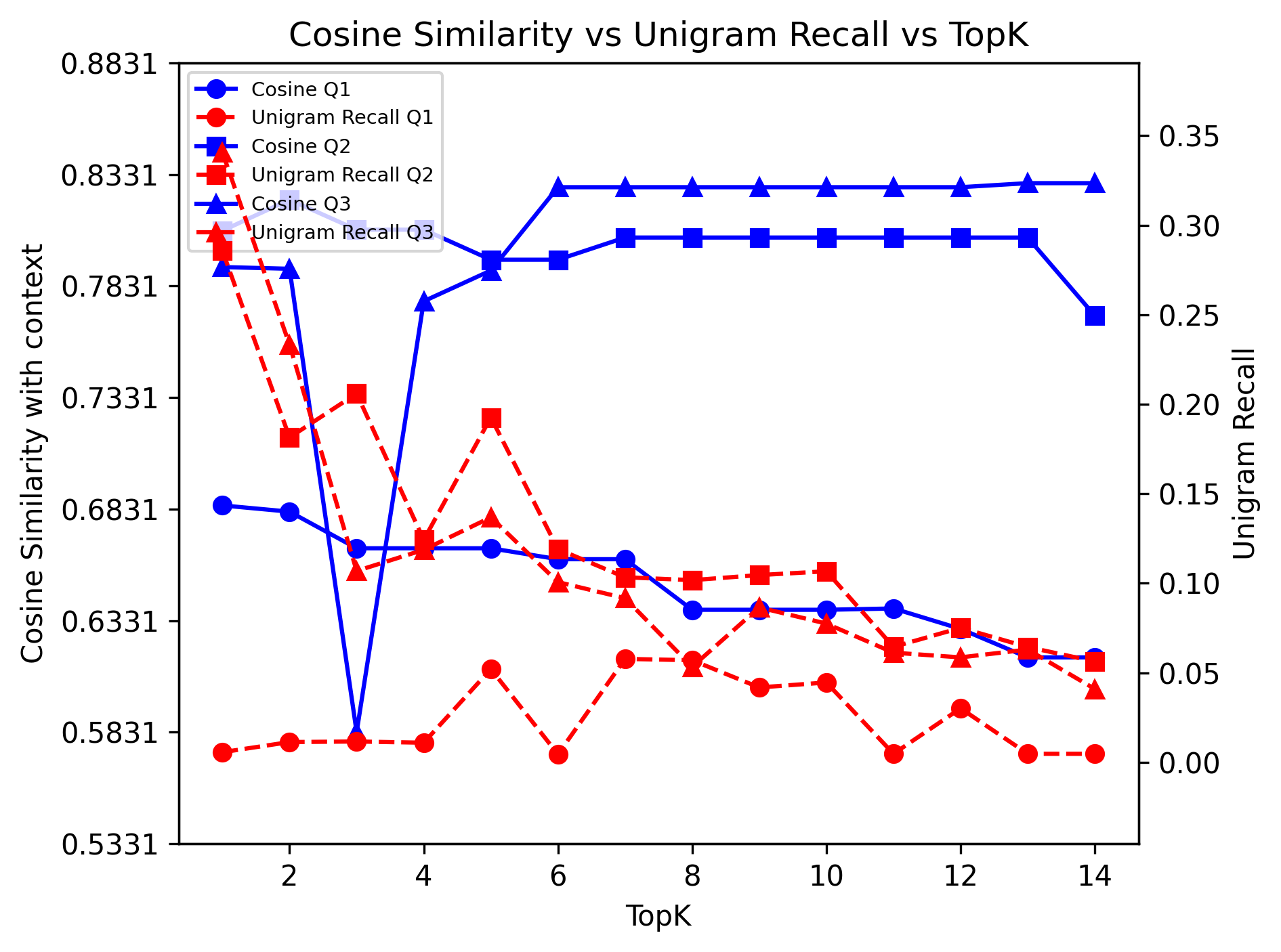}
            \caption{Llama3: Factual Dense}
            \label{fig:llama_factual_dense}
        \end{subfigure}
        \hfill
        \begin{subfigure}{0.48\textwidth}
            \includegraphics[width=\linewidth]{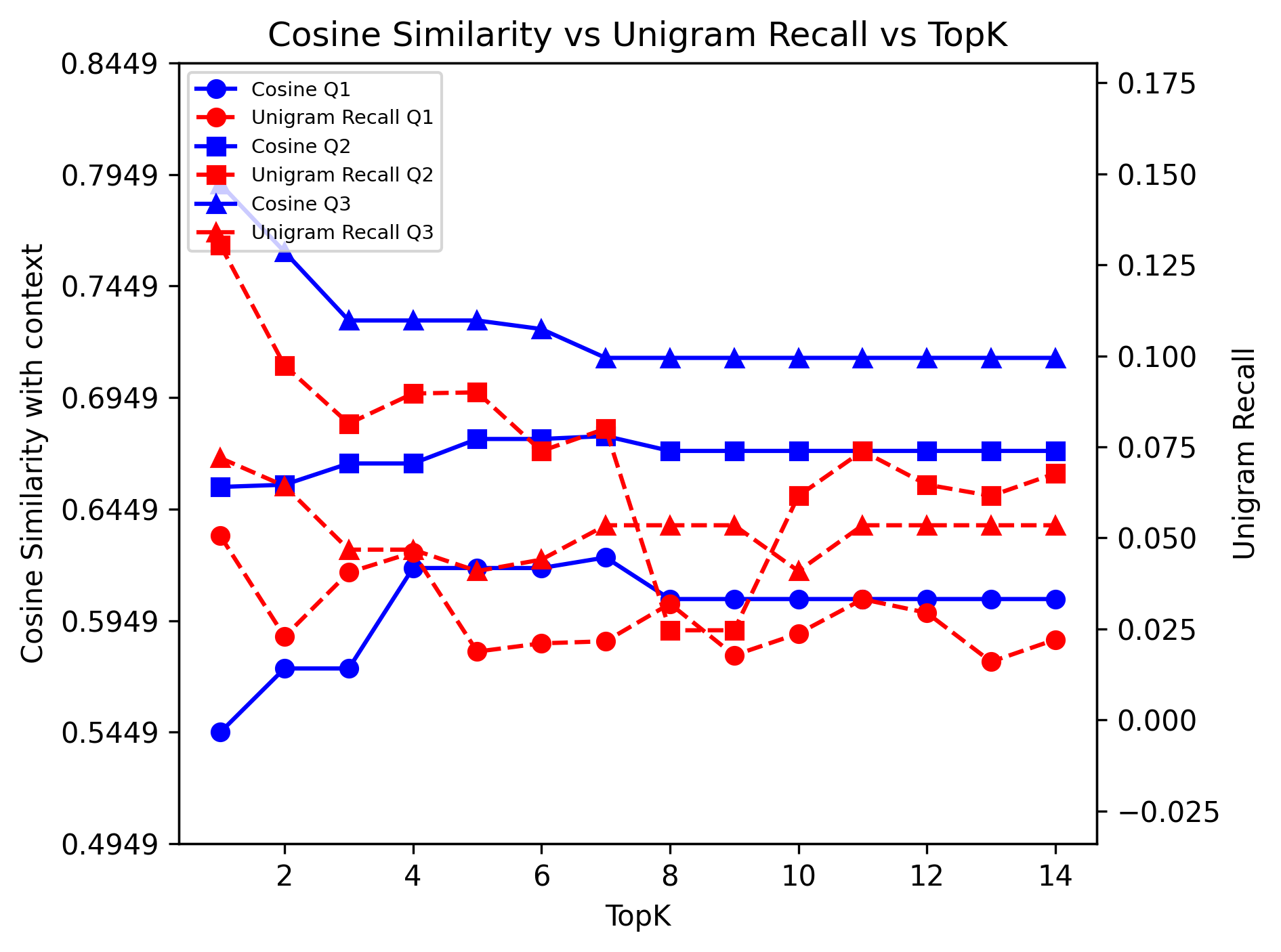}
            \caption{Llama3: Factual Sparse}
            \label{fig:llama_factual_sparse}
        \end{subfigure}
    \end{minipage}
    \caption{Analysis of Cosine Similarity and Unigram Recall vs TopK for Mistral8x7B LLM model (left) and Llama3-8B LLM model (right)}
    \label{fig:combined_plots}
\end{figure*}

\subsubsection{Unigram Recall} Unigram recall is the fraction of words in the reference/context that also appear in the LLM generated answer. 
We can observe from Figure\ref{fig:combined_plots_2} that in general recall does not increase even when we increase top-k value. This can be explained by the fact that for this enterprise specific data, adding more chunks as context to LLM in the prompt do not necessarily help improve response's recall because the text in additional chunks is not relevant to the query. This also correlates with the observation of no improvement in cosine similarity of retrieved context with the query as top-k value is increased. Based on this one can infer that there is no defined correlation between cosine similarity and Unigram recall so this metric is not useful in generating a good measure of answer quality.

\begin{figure*}[ht]
    \centering
    \begin{minipage}{0.49\textwidth} % Adjusted width of the minipage
        \centering
        \begin{subfigure}{0.48\textwidth} % Adjusted width of subfigures
            \includegraphics[width=\linewidth]{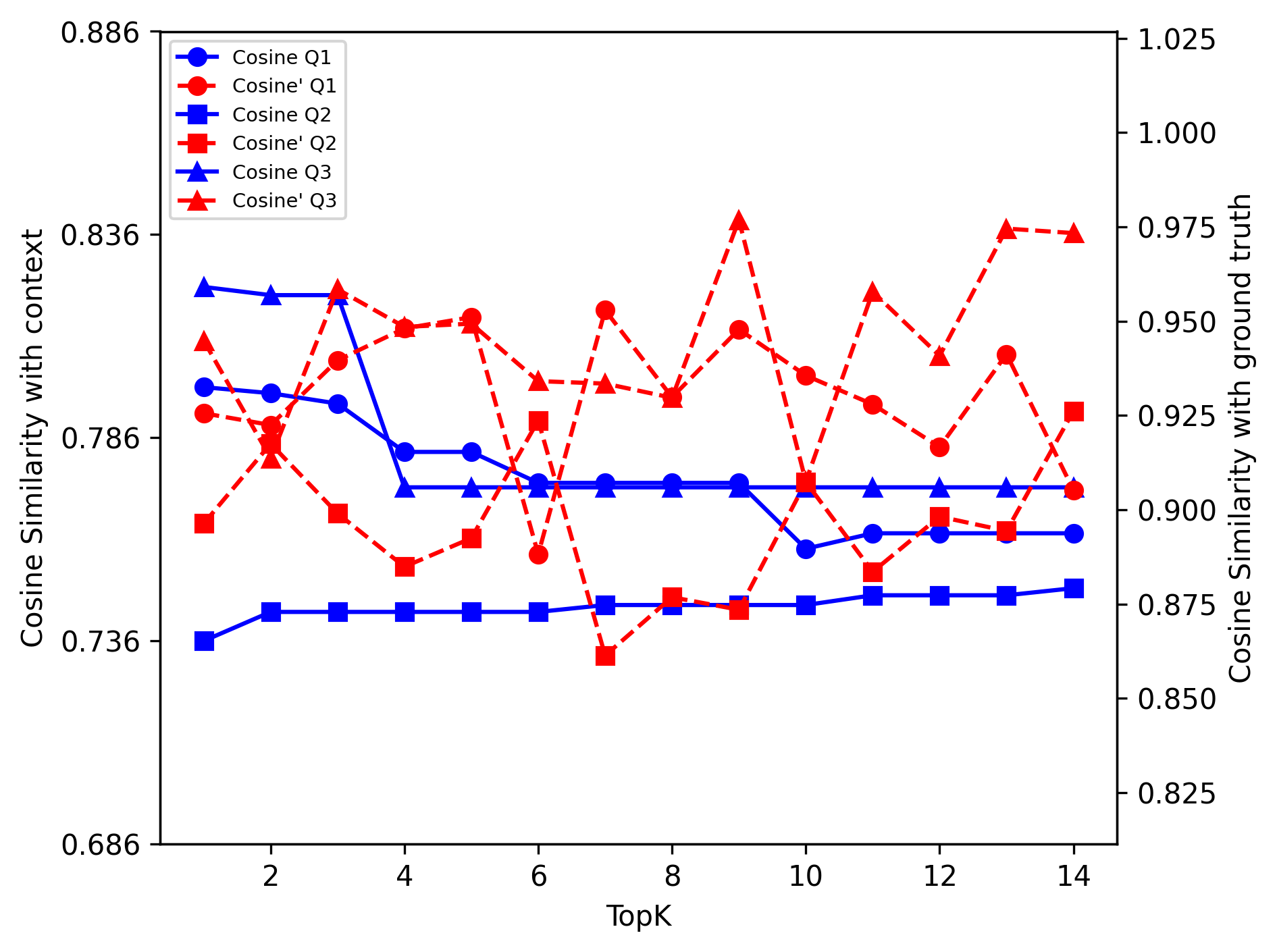}
            \caption{Mistral: Reason Dense}
            \label{fig:mistral_reson_dense_cosine}
        \end{subfigure}
        \hfill
        \begin{subfigure}{0.48\textwidth}
            \includegraphics[width=\linewidth]{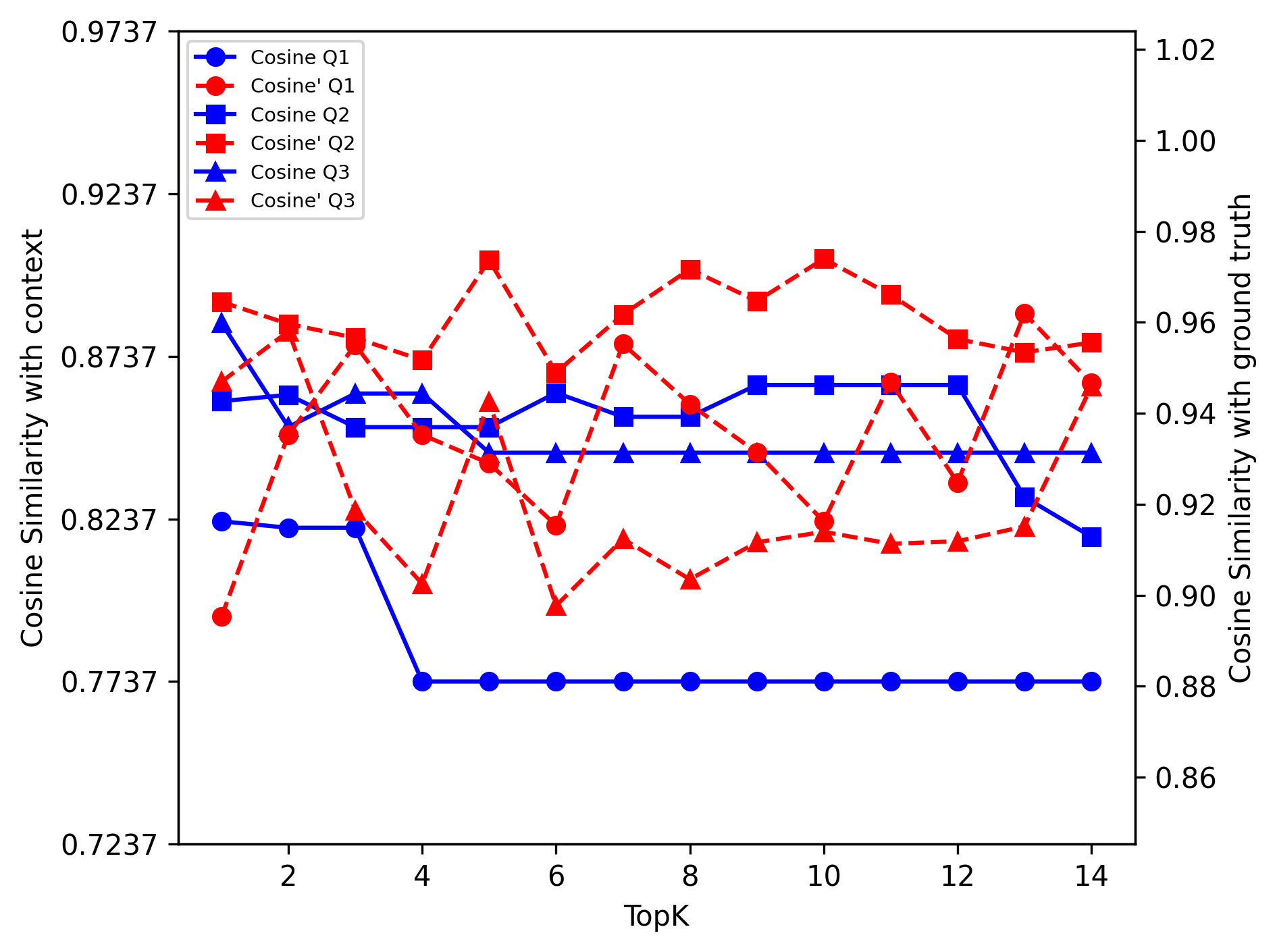}
            \caption{Mistral: Reason Sparse}
            \label{fig:mistral_factual_dense_factual_sparse_cosine}
        \end{subfigure}
        \vspace{0.2cm} % Adjust vertical space between rows
        \begin{subfigure}{0.48\textwidth}
            \includegraphics[width=\linewidth]{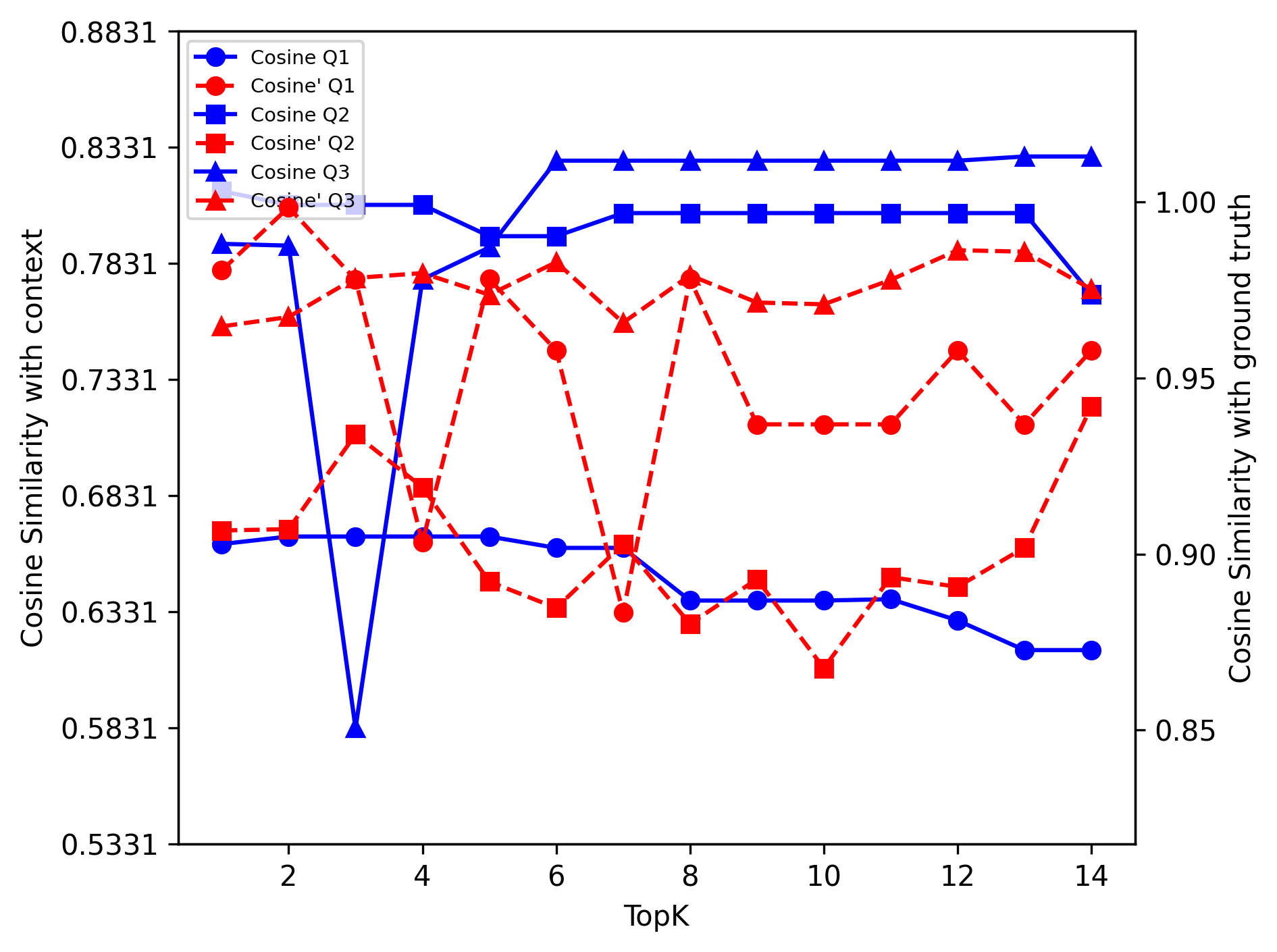}
            \caption{Mistral: Factual Dense}
            \label{fig:mistral_reson_dense_reason_sparse_trendlines_cosine}
        \end{subfigure}
        \hfill
        \begin{subfigure}{0.48\textwidth}
            \includegraphics[width=\linewidth]{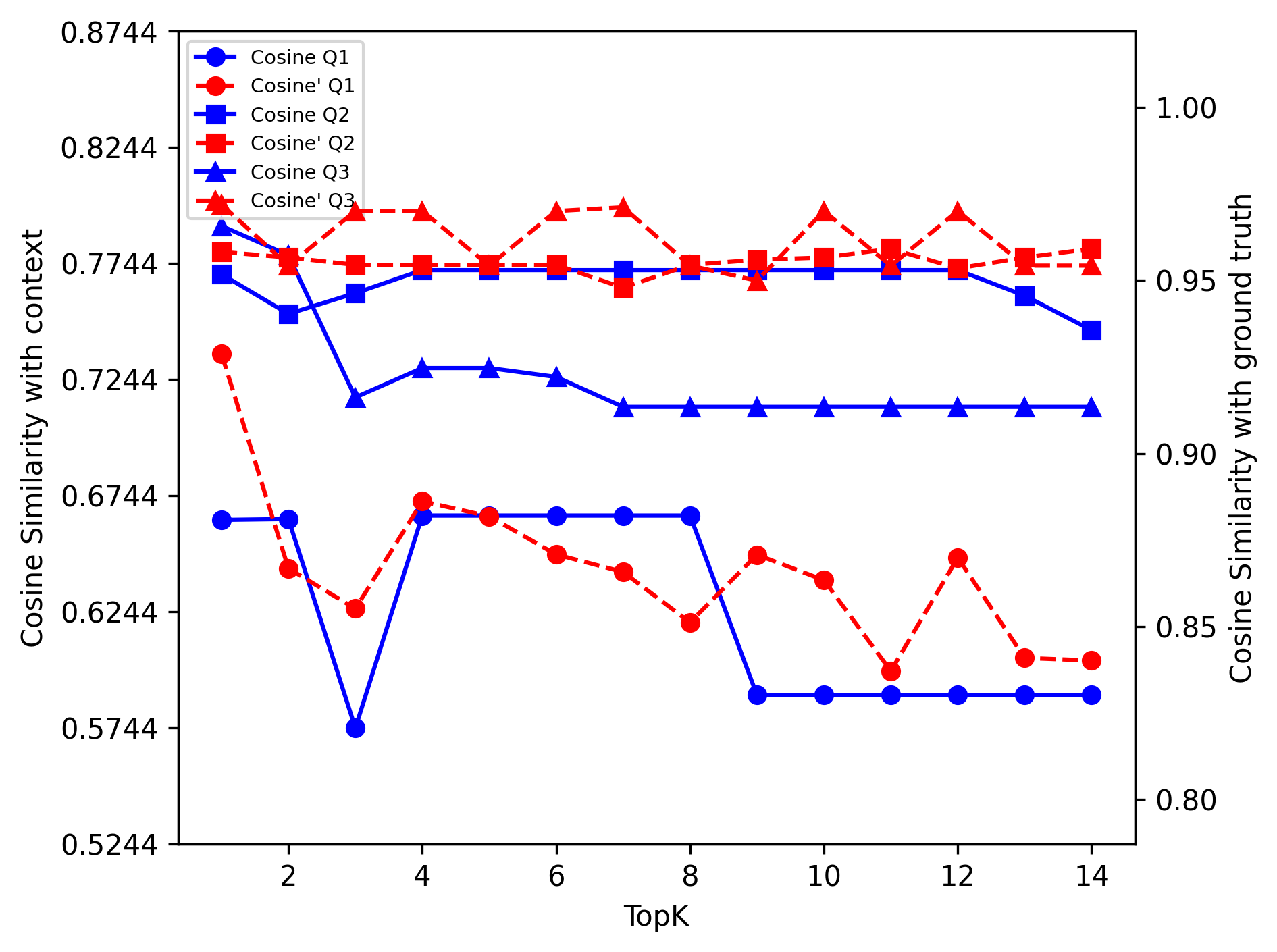}
            \caption{Mistral: Factual Sparse}
            \label{fig:mistral_factual_dense_factual_sparse_trendlines_cosinee}
        \end{subfigure}
    \end{minipage}
    \hfill
    \begin{minipage}{0.48\textwidth} % Adjusted width of the minipage
        \centering
        \begin{subfigure}{0.48\textwidth} % Adjusted width of subfigures
            \includegraphics[width=\linewidth]{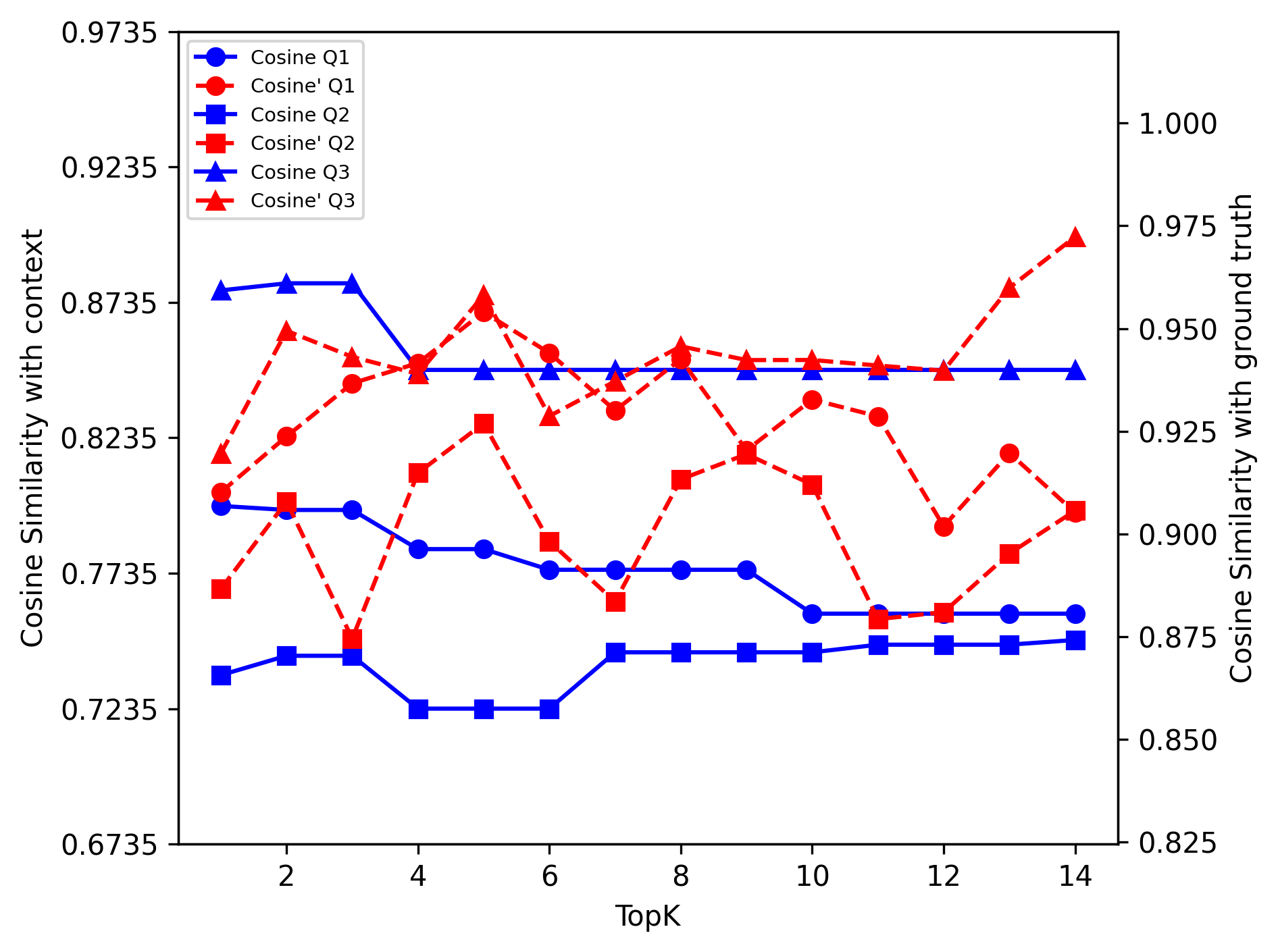}
            \caption{LLama3: Reason Dense}
            \label{fig:llama_reson_dense_reason_sparse_cosine}
        \end{subfigure}
        \hfill
        \begin{subfigure}{0.48\textwidth}
            \includegraphics[width=\linewidth]{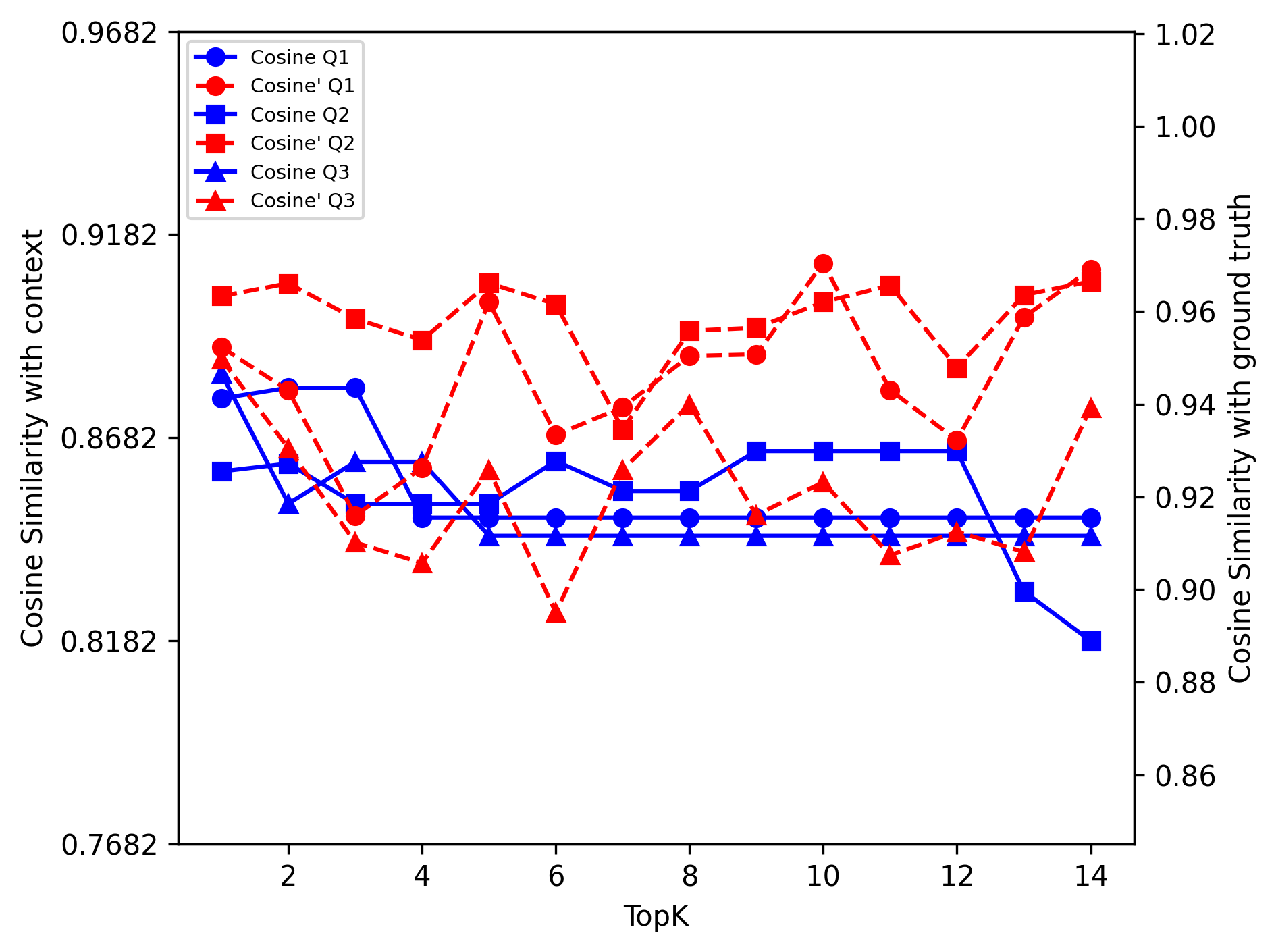}
            \caption{LLama3: Reason Sparse}
            \label{fig:llama_factual_dense_factual_sparse_cosine}
        \end{subfigure}
        \vspace{0.1cm} % Adjust vertical space between rows
        \begin{subfigure}{0.48\textwidth}
            \includegraphics[width=\linewidth]{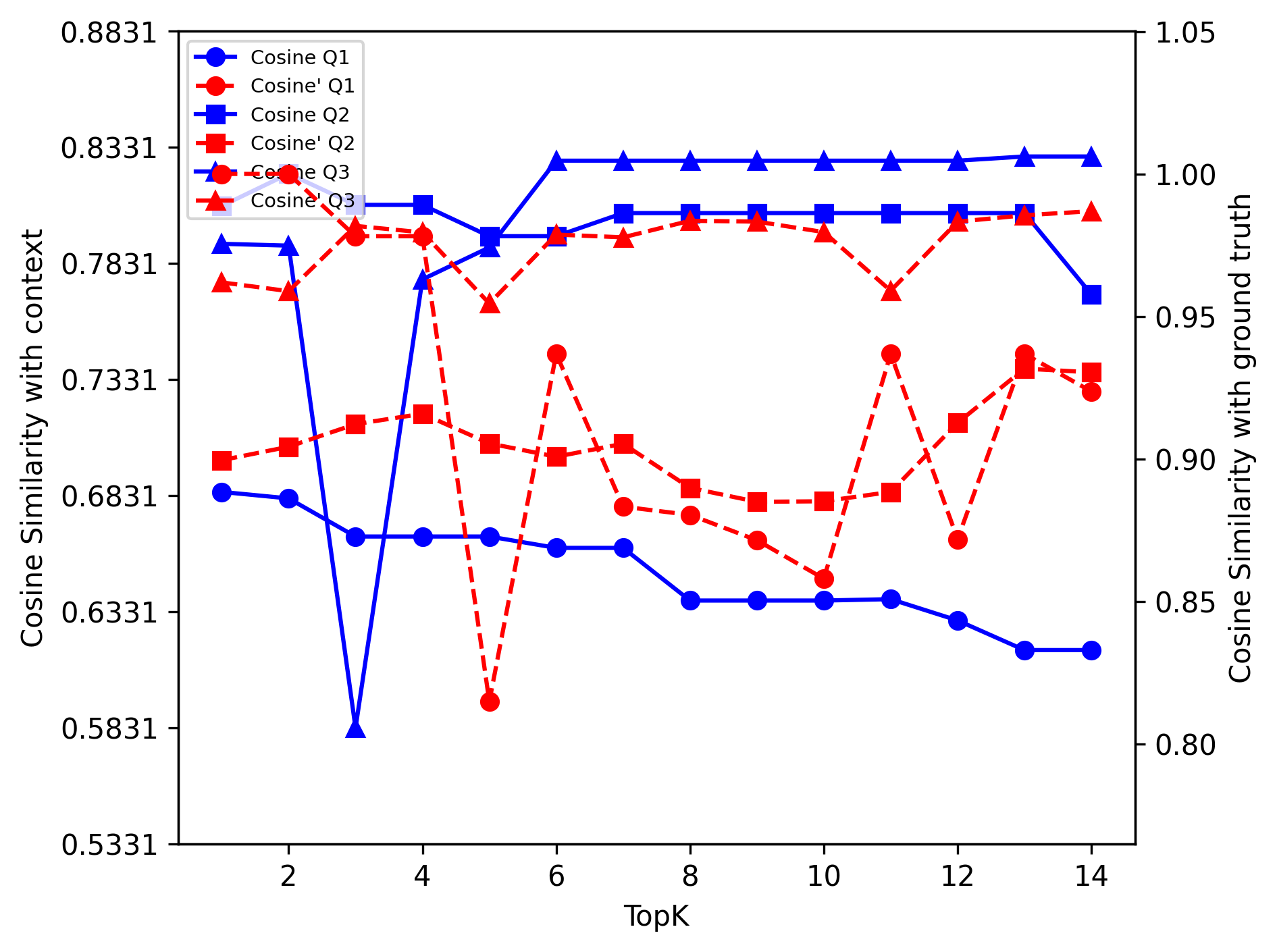}
            \caption{LLama3: Factual Dense}
            \label{fig:llama_reson_dense_reason_sparse_trendlines_cosine}
        \end{subfigure}
        \hfill
        \begin{subfigure}{0.48\textwidth}
            \includegraphics[width=\linewidth]{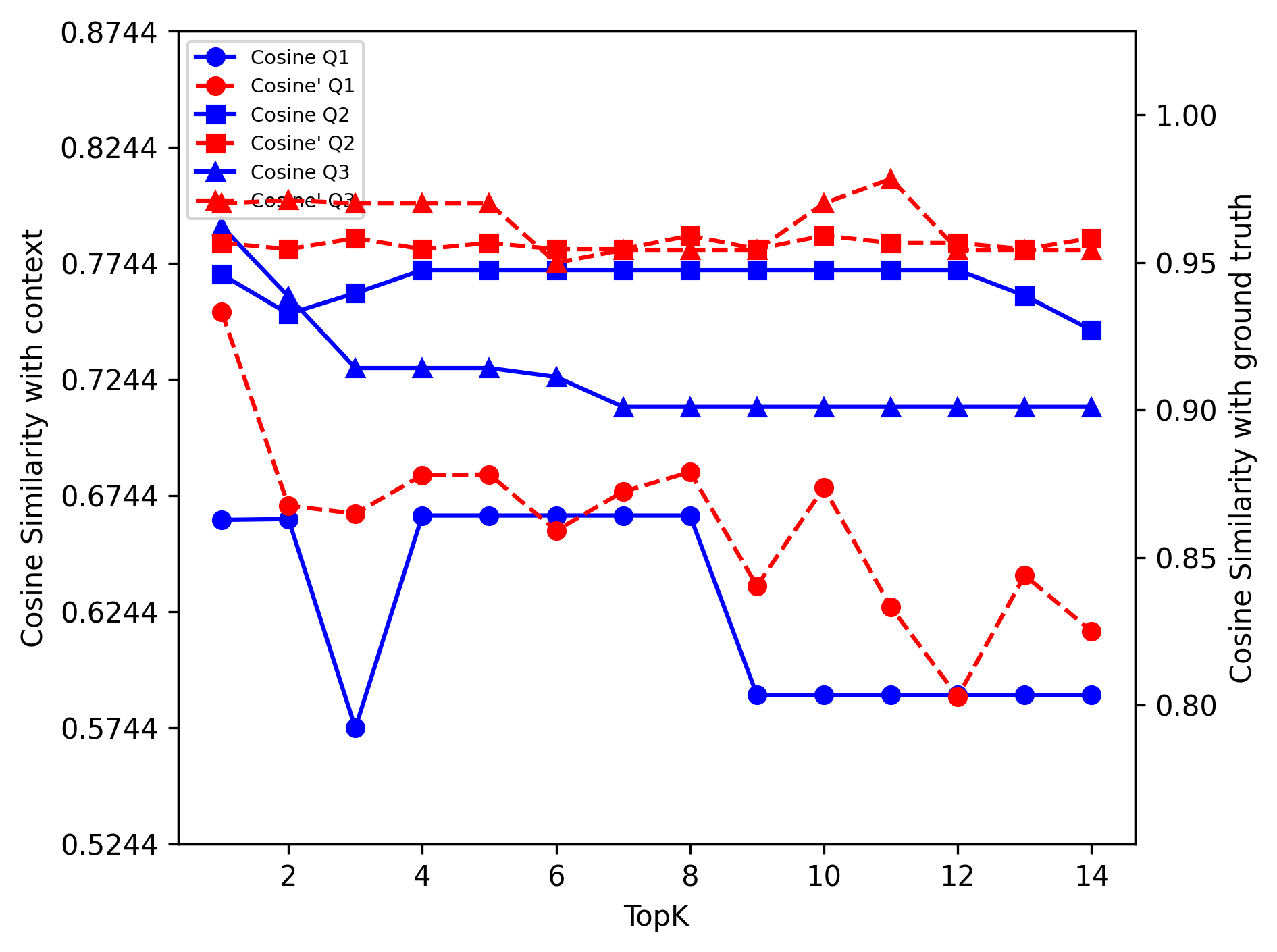}
            \caption{LLama3: Factual Sparse}
            \label{fig:llama_factual_dense_factual_sparse_trendlines_cosine}
        \end{subfigure}
    \end{minipage}
    \caption{Analysis of Cosine Similarity with context vs Cosine similarity with ground truth vs TopK for Mistral8x7B LLM model (left) and LLama3-8B LLM model (right)}
    \label{fig:combined_plots_3}
\end{figure*}

\subsubsection{Cosine similarity with ground truth answer(CSGA)} This metric measures the cosine similarity between the LLM generated answer and the ground truth answer generated using GPT-4. From Figure\ref{fig:combined_plots_3}we can infer that the values more or less remains constant as top-k increases signifying that the retrieved answer does  not  improve a lot even with increased top-k as the extra chunks will become irrelevant to the query.

In our analysis of Mistral and Llama3, we observe the distinct patterns in performance metrics across various question types. For reason dense questions, both models exhibited consistent performance values, with the exception of a single question where a increase was noted. This suggests that both models generally handle reason dense questions with stable accuracy, barring occasional outliers.In contrast, for reason sparse and factual dense questions, the performance values remained uniformly constant across all six questions for both Mistral and Llama3. This indicates a robust and consistent handling of reason sparse and a general reliability in processing factual dense questions with the exception of few anomalies. Lastly, for factual sparse questions, there is a noticeable decrease in performance values for one question, while the other two questions showed consistent values. This suggests that factual sparse questions may occasionally challenge the models, resulting in variable performance on certain questions. We further analysed the performance of Mistral and Llama3 using Deepeval framework. 
\newpage
\newpage
\subsection{Evaluation using deepeval scores}
For this evaluation NLTKTextSplitter has been used. Table\ref{tab:llama} provides the evaluation done for Llama3-8B and based on BAAI/bge-large-en-v1.5 embedding. 
\begin{table}[H]
    \centering
    \caption{Performance Metrics for Llama3-8B}
    \label{tab:llama}
    \resizebox{\columnwidth}{!}{
    \begin{tabular}{lcccc}
        \toprule
        Metrics & Reason Dense & Reason Sparse & Factual Dense & Factual Sparse \\ \midrule
        \textbf{Unigram Precision} & & & & \\ \midrule
        Average & 0.737 & 0.789 & 0.81 & 0.82 \\ 
        Median & 0.709 & 0.774 & 0.8 & 0.81 \\ \midrule
        \textbf{Contextual Precision} & & & & \\ \midrule
        Average & 0.911 & 0.938 & 0.864 & 0.85 \\ \midrule
        %Median & 1 & 1 & 1 & 1 \\ \midrule
        \textbf{Contextual Recall} & & & & \\ \midrule
        Average & 0.92 & 0.98 & 0.916 & 0.92 \\ \midrule
        %Median & 1 & 1 & 1 & 1 \\ \midrule
        \textbf{Contextual Relevancy} & & & & \\ \midrule
        Average & 0.68 & 0.6363 & 0.66 & 0.947 \\ 
        Median & 1 & 1 & 1 & 1 \\ \midrule
        \textbf{Answer Relevancy} & & & & \\ \midrule
        Average & 0.98 & 0.93 & 0.97 & 1 \\ \midrule
        %Median & 1 & 1 & 1 & 1 \\ \midrule
        \textbf{Time (s)} & & & & \\ \midrule
        Average & 2.88 & 2.732 & 1.6004 & 1.74 \\ 
        Median & 2.61 & 2.25 & 1.472 & 1.401 \\ \midrule
        \textbf{CSGA range} & & & & \\ \midrule
        Range & [0.875,0.975] & [0.87,0.97] & [0.81,1]&[0.8,0.98] \\
        \bottomrule
    \end{tabular}
    }
\end{table}

The same has been done for mistral 8x7B LLM based on BAAI/bge-large-en-v1.5 embedding model, refer Table\ref{tab:mistral}.
\begin{table}[H]
    \centering
    \caption{Performance Metrics for Mistral 8x7B}
    \label{tab:mistral}
    \resizebox{\columnwidth}{!}{
    \begin{tabular}{lcccc}
        \toprule
        Metrics & Reason Dense & Reason Sparse & Factual Dense & Factual Sparse \\ \midrule
        \textbf{Unigram Precision} & & & & \\ \midrule
        Average & 0.69 & 0.67 & 0.73 & 0.73 \\ 
        Median & 0.74 & 0.72 & 0.754 & 0.756 \\ \midrule
        \textbf{Contextual Precision} & & & & \\ \midrule
        Average & 0.9 & 0.94 & 0.8 & 0.85 \\ \midrule
        %Median & 1 & 1 & 1 & 1 \\ \midrule
        \textbf{Contextual Recall} & & & & \\ \midrule
        Average & 0.94 & 0.98 & 0.93 & 0.91 \\ \midrule
        %Median & 1 & 1 & 1 & 1 \\ \midrule
        \textbf{Contextual Relevancy} & & & & \\ \midrule
        Average & 0.6 & 0.56 & 0.73 & 0.92 \\ 
        Median & 1 & 0 & 1 & 1 \\ \midrule
        \textbf{Answer Relevancy} & & & & \\ \midrule
        Average & 0.94 & 0.87 & 0.95 & 0.91 \\ \midrule
        %Median & 1 & 1 & 1 & 1 \\ \midrule
        \textbf{Time} & & & & \\ \midrule
        Average & 2.73 & 2.91 & 1.8 & 1.6 \\ 
        Median & 2.96 & 3.06 & 1.4 & 1.36 \\ \midrule
        \textbf{CSGA range} & & & & \\ \midrule
        Range & [0.87,0.975] & [0.89,0.98] & [0.84,1]&[0.84,0.98] \\
        \bottomrule
    \end{tabular}
    }
\end{table}

We observe that Llama3-8B significantly outperformed Mistral8x7B in all tasks in terms of unigram precision, indicating that the Llama3-8B model is highly effective at capturing relevant information at a granular level. Additionally, the inference time for Llama3-8B is on par with or better than Mistral8x7B, as shown in Table \ref{tab:llama} and Table \ref{tab:mistral}. The variations in contextual metrics suggest that the performance of both models can be task-dependent. Mistral performs significantly worse on sparse reasoning and  dense factual information retrieval, while it is comparable on dense reasoning and  sparse factual information tasks. Despite having fewer parameters (8 billion for Llama3-8B) compared to Mistral8x7B(56 Billion), Llama3-8B has a slight edge over its competitor. Its ability to maintain high precision, coupled with lower inference times, makes it a robust option for text generation using enterprise specific data. These findings suggest that the Llama3-8B model can serve as a more efficient and effective alternative across all 4 question categories. 

Moreover, comparing these open-source models such as Llama3-8B and Mistral8x7B with GPT-3.5 indicate that they outperform GPT-3.5 in several key areas:

\begin{itemize}
    \item \textbf{Unigram Precision}: While GPT-3.5 achieves a unigram precision of 0.77, Llama3-8B slightly outperforms it ranging as values from 0.737 to 0.82 , while mistral 8x7B under performs with values ranging from  0.67 to 0.73 respectively.
    \item \textbf{Contextual Recall}: Both Llama3-8B (0.916-0.98) and Mistral (0.91-0.98) surpass GPT-3.5 (0.86) in contextual recall, demonstrating their superior ability to retrieve relevant context.
    \item \textbf{Contextual Relevancy}: Llama3-8B (0.636-0.947) and Mistral (0.56-0.92) significantly outperform GPT-3.5 (0.60) in contextual relevancy, indicating better alignment with the intended context.
    \item \textbf{Answer Relevancy}: Llama3-8B (0.93-1.00) performs on par with GPT-3.5 (1), with Mistral (0.87-0.95) shows under performed results.
    \item \textbf{Contextual Precision}: Despite these strengths, Llama3-8B (0.85-0.938) and Mistral (0.8-0.94) fall short in contextual precision compared to GPT-3.5 (0.98).
\end{itemize}

These findings suggest that open-source models like Llama3-8B and Mistral offer notable improvements over GPT-3.5, particularly in contextual recall and relevancy,answer relevancy and unigram precision, though they may still have some limitations in contextual precision\cite{cos}.

\section{Discussion}
As highlighted in Section \ref{sec:Result}, the Llama3-8B model demonstrates superior performance compared to the Mistral 8x7B model across various tasks. This advantage is likely attributed to Llama3-8B's training on an extensive and diverse dataset encompassing over 15 trillion tokens. Additionally, the Llama3-8B model benefits from instruction-tuning, a process that optimizes it for tasks requiring adherence to user instructions, thereby enhancing its effectiveness for RAG-based applications. It is crucial to note that the results presented in this study are specific to the datasets used and should not be generalized to other datasets. Furthermore, from a cost-efficiency standpoint, these open-source alternatives offer significant cost reductions.

From the top-k vs. cosine similarities graph, it can be inferred that beyond a certain top-k value, the retrieved information becomes increasingly irrelevant to the query. This irrelevance leads to a plateau in the graph across all question sets for both models, indicating that additional documents do not contribute to the query's answer.

For reasoning-dense questions, the Llama3 model effectively utilized the retrieved context better than the Mistral model, suggesting that models with a smaller context window can sometimes use information more efficiently than those with a larger context window. However, for factually dense questions, the Mistral model benefited from the additional retrieved chunks, showing a steady increase in cosine similarity for two of the questions which can be attributed the mixture of experts being a effective way here.

In reasoning-sparse questions, the Llama3 model consistently utilized the retrieved information more effectively than the Mistral model, which showed significant variations across different top-k values. This variation indicates that Mistral did not utilize the retrieved information as efficiently, although both models eventually achieved similar precision scores.

For factually sparse questions, both models performed similarly as top-k increased for two questions, with a notable increase for one question. This result suggests that for factually sparse questions, both models were able to utilize the retrieved information to a comparable extent.

To compare the quality of answers generated by an open-source LLM and GPT-4, as we can infer from Figure\ref{fig:combined_plots_2} cosine similarity proves to be a reliable metric. This metric remains relatively stable even with variations in the top-k parameter, exhibiting minimal changes (typically around 0.5 and a maximum of 1) across different questions. In contrast, unigram recall and precision, which measure the proportion of single words correctly matched between the generated and reference answers, show greater variability with average changes near 1 and maximum changes up to 2. This variability suggests that unigram recall and precision are less reliable for assessing answer quality compared to the consistent performance of cosine similarity with ground truth answer.

As we can infer from Table \ref{tab:llama} and Table \ref{tab:mistral}, the values of  Cosine similarity with the groundtruth answer (CSGA) do not vary much across all categories of question. Also, these values compare well with Contextual Recall, Contextual Precision and Answer relevancy metrics of Deepeval framework. This consistency makes CSGA a reliable metric. Besides, cosine similarly is easier to calculate compared to Deepeval , which utilizes four evaluation metrics and requires significant inference time for every question. Additionally, we can also infer from appendix that the average time taken for Llama3 and Mistral using perplexity API is nearly 50\% lower compared that of GPT-3.5 .

\section{Conclusion}
This work investigates the efficacy of open source LLMs in providing response to questions related to enterprise specific data. For the same, vector database using open source embedding has been created followed by categorising the questions into 4 categories viz. sparse factual, sparse reason, dense factual and dense reason. Here are some salient findings from this investigation:

\sloppy
\begin{itemize}
    \item \textbf{Effectiveness of Open-Source LLMs in RAG Systems}: The study demonstrates that open-source LLMs integrated within Retrieval-Augmented Generation (RAG) framework,  generate response of similar accuracy and relevance of as commercial LLMs.
    
    \item \textbf{Influence of context length}: RAG based QA evaluation by increasing provided context (by varying top-k) demonstrates no significant improvement in answer quality as CSGA does not change much . Thus, this evaluation on enterprise specific data shows that one need not  have very large LLM context window for  Question Answering (QA) task. 
    %%\item \textbf{Comprehensive Evaluation Metrics}: Utilizing ROUGE scores alongside DeepEval contextual metrics provides a thorough evaluation of the RAG system's performance, capturing both quantitative and qualitative aspects of the generated responses. This multifaceted approach ensures a robust assessment of the system's capabilities. ( can be removed) 

    \item \textbf{Llama3-8B vs mistral8x7B}: LLM parameter count need not necessarily improve RAG based Question Answering (QA) , as evident from Llama3 outperforming Mistral. 
    
    Especially for proprietary enterprise datasets as it's difficult for RAG based systems to 
    perform on them over regular open source data sets available online.
    
    \item \textbf{Performance}: The use of open-source LLMs to build RAG based QA system provides performance similar to commercial LLMs. Besides, this work also demonstrates that open source LLMs  can be scaled and adapted to enterprise-specific data sets without need of investing in expensive Graphical Processing Unit(GPU)s for real time inferencing.
% available through API providers like Perplexity
    \item \textbf{Cosine Similarity with Groundtruth answer (CSGA)  }:  is an effective metric for measuring answer quality of RAG answers. 
\end{itemize}

\section*{Acknowledgement}
The authors thank  I-Venture at Indian School of Business for infrastructural support toward this work. Authors are extremely grateful to Prof. Bhagwan Chowdhry, Faculty Director (I-Venture at ISB) and Rahul Sundar(Scientist at Verisk) for his continued encouragement and support to carry out this research.

%%
%% The acknowledgments section is defined using the "acks" environment
%% (and NOT an unnumbered section). This ensures the proper
%% identification of the section in the article metadata, and the
%% consistent spelling of the heading.
%%
%% Print the bibliography
%%
\printbibliography

@article{lewis2020retrieval,
	title        = {Retrieval-Augmented Generation for Knowledge-Intensive NLP Tasks},
	author       = {Lewis, Patrick and Perez, Ethan and Piktus, Aleksandra and Petroni, Fabio and Karpukhin, Vladimir and Goyal, Naman and Küttler, Heinrich and Lewis, Mike and Yih, Wen-tau and Rockt{\"a}schel, Tim and Riedel, Sebastian and Kiela, Douwe},
	year         = 2020,
	journal      = {arXiv preprint arXiv:2005.11401}
}

@article{karpukhin2020dense,
	title        = {Dense Passage Retrieval for Open-Domain Question Answering},
	author       = {Karpukhin, Vladimir and Oguz, Barlas and Min, Sewon and Lewis, Patrick and Wu, Ledell and Edunov, Sergey and Chen, Danqi and Yih, Wen-tau},
	year         = 2020,
	journal      = {arXiv preprint arXiv:2004.04906},
	url          = {https://arxiv.org/pdf/2004.04906}
}

@article{purwar2024keyword,
	title        = {Keyword Augmented Retrieval: Novel framework for Information Retrieval integrated with speech interface},
	author       = {Anupam Purwar and Rahul Sundar},
	year         = 2023,
	journal      = {arXiv preprint arXiv:2310.04205},
	url          = {https://dl.acm.org/doi/10.1145/3639856.3639916}
}

@article{cost_analysis2024,
	title        = {A Cost Analysis of Generative Language Models and Influence Operations},
	author       = {Micah Musser},
	year         = 2023,
	journal      = {arXiv preprint arXiv:2308.03740},
	url          = {https://arxiv.org/pdf/2308.03740}
}

@article{Afinetune,
	title        = {A FINE-TUNING ENHANCED RAG SYSTEM WITH QUANTIZED INFLUENCE MEASURE AS AI JUDGE},
	author       = {Keshav Rangan and Yiqiao Yin},
	year         = 2024,
	journal      = {arXiv preprint arXiv::2402.17081v1},
	url          = {https://arxiv.org/pdf/2402.17081v1}
}

@article{UL2,
	title        = {UL2: Unifying Language Learning Paradigms},
	author       = {Yi Tay and Mostafa Dehghani and Vinh Q. Tran and  Xavier Garcia and Jason Wei and Xuezhi Wang and Hyung Won Chung and Siamak Shakeri and Dara Bahri and Tal Schuster and Huaixiu Steven Zheng and  Denny Zhou and Neil Houlsby and Donald Metzler},
	year         = 2023,
	journal      = {arXiv preprint arXiv::2205.05131},
	urll         = {https://arxiv.org/pdf/2205.05131}
}

@article{COLB,
	title        = {ColBERT: Efficient and Effective Passage Search via Contextualized Late Interaction over BERT},
	author       = {Omar Khattab and Matei Zaharia},
	year         = 2020,
	journal      = {arXiv preprint arXiv::2004.12832},
	url          = {https://arxiv.org/pdf/2004.12832}
}

@article{Eval,
	title        = {Evaluation of Retrieval-Augmented Generation: A Survey},
	author       = {Hao Yu and Aoran Gan and Kai Zhang and Shiwei Tong and  Qi Liu and Zhaofeng Liu},
	year         = 2024,
	journal      = {arXiv preprint arXiv::2405.07437v1},
	url          = {https://arxiv.org/pdf/2405.07437v1}
}

@article{RAGAS,
	title        = {RAGAS: Automated Evaluation of Retrieval Augmented Generation},
	author       = {Shahul Es and Jithin James and Luis Espinosa-Anke and Steven Schockaert},
	year         = 2023,
	journal      = {arXiv preprint arXiv::2309.15217},
	url          = {https://arxiv.org/pdf/2309.15217}
}

@article{MTEB,
	title        = {MTEB: Massive Text Embedding Benchmark},
	author       = {Niklas Muennighoff1 and Nouamane Tazi1 andLoïc Magne1 and Nils Reimers},
	year         = 2023,
	journal      = {arXiv preprint arXiv::2210.07316},
	url          = {https://arxiv.org/pdf/2210.07316}
}

@misc{langchain,
	year         = 2024,
	url          = {https://python.langchain.com/v0.1/docs/get_started/introduction},
	note         = {langchain documentation}
}

@misc{huggingface,
	year         = 2024,
	url          = {https://huggingface.co/models},
	note         = {huggingface documentation}
}

@misc{deepeval,
	year         = 2024,
	url          = {https://docs.confident-ai.com/},
	note         = {deepeval documentation}
}

@misc{perp,
	year         = 2024,
	url          = {https://docs.perplexity.ai/docs/getting-started},
	note         = {perplexity api documentation}
}

@article{cos,
	title        = {COS-Mix: Cosine Similarity and Distance Fusion for Improved Information Retrieval},
	author       = {Kush Juvekar and Anupam Purwar},
	year         = 2024,
	url          = {https://arxiv.org/pdf/2406.00638}
}

@article{llmllm,
    author = {Lianmin Zheng and Wei-Lin Chiang and Ying Sheng and Siyuan Zhuang and Zhanghao Wu and Yonghao Zhuang and Zi Lin and Zhuohan Li and Dacheng Li and Eric P. Xing and Hao Zhang and Joseph E. Gonzalez and Ion Stoica},
    title = {Judging LLM-as-a-Judge with MT-Bench and Chatbot Arena},
    url = {https://arxiv.org/pdf/2306.05685},
    year = {2024}
}

@misc{nltk,
    year=2024 ,
    url={https://api.python.langchain.com/en/latest/nltk/langchain_text_splitters.nltk.NLTKTextSplitter.html},
    note = {NLTKsplitter documentation}
}

@misc{faiss,
    year=2024 ,
    note = {FAISS documentation},
    url={https://python.langchain.com/v0.2/docs/integrations/vectorstores/faiss/}
}

@misc{BAAI,
    year=2023,
    note ={BAAI embedding documentation},
    url={https://huggingface.co/BAAI/bge-large-en-v1.5},
}

@misc{BM25,
    url={https://python.langchain.com/v0.2/docs/integrations/retrievers/bm25/} ,
   year=2024 ,
    note={BM25 documentation}
}

@misc{cost,
url={https://openai.com/api/pricing/},note={OpenAI API pricing documentation},year=2024}

@misc{perpcost,
    url={https://docs.perplexity.ai/docs/pricing},
    year=2024,
    note={Perplexity pricing documentation}
}

@misc{neur,
    url={https://docs.neuralmagic.com/get-started/finetune/} ,
    year=2024,
    note={neuralmagic documentation}
}

@misc{scal,
    url={https://blog.lancedb.com/optimizing-llms-a-step-by-step-guide-to-fine-tuning-with-peft-and-qlora-22eddd13d25b/},
    year=2024,
    note={lancedb documentation}
}

@misc{recr,
    year=2024 ,
    url={https://python.langchain.com/v0.1/docs/modules/data_connection/document_transformers/recursive_text_splitter/},
    note = {RecursiveCharacterTextSplitter langchain documentation}
}

@misc{bge_embedding,
      title={C-Pack: Packaged Resources To Advance General Chinese Embedding}, 
      author={Shitao Xiao and Zheng Liu and Peitian Zhang and Niklas Muennighoff},
      year={2023},
      eprint={2309.07597},
      archivePrefix={arXiv},
      primaryClass={cs.CL}
}

@misc{wrapper,
    url={https://github.com/amaze18/dlabs_hybrid_search/blob/main/pplx.py},
    note = {Perplexity wrapper code},
    year=2024
}

@misc{perplex,
    url={https://docs.perplexity.ai/docs/model-cards} ,
    year=2024,
    note={Perplexity models documentation}
}
%%
%% If your work has an appendix, this is the place to put it.
\appendix
\section*{Appendix}
\begin{figure}[h]
    \centering
    \includegraphics[width=0.5\linewidth]{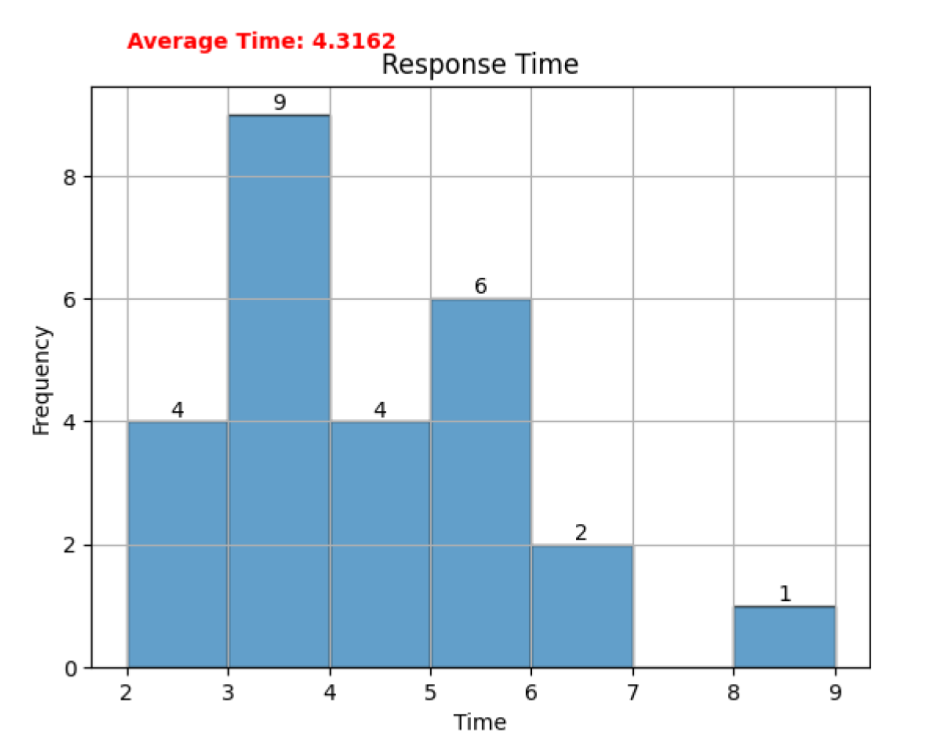} % Adjust width as needed
    \caption*{Histogram of inference time using GPT 3.5: Average response time 4.3 seconds}
    \label{hist}
\end{figure}

\end{document}